# Commissioning of SLAC SLD 45-Degree Chambers [*]

V.O. Eschenburg

Stanford Linear Accelerator Center
Stanford University
Stanford, CA 94309

---

[*] M.S.. thesis, Stanford Linear Accelerator Center OR Stanford University, Stanford, CA 94309.

# Commissioning of SLAC SLD 45-Degree Chambers

BY

Vance Onno Eschenburg

Approved:

___________________________________    ___________________________________
Dr. Robert Kroeger                     Dr. Jim Reidy
Director of the Thesis

___________________________________
Dr. Don Summers

___________________________________
Dean of the Graduate School            December 2001

Commissioning of SLAC SLD 45-Degree Chambers

By

Vance Onno Eschenburg

B.A. University of California at Berkeley, 1995

A Thesis
Submitted to the Faculty of
The University of Mississippi
For the Degree of Master of Science
In the Department of Physics

The University of Mississippi
December 2001

Dedication

This thesis is dedicated to my parents,

Captain Vernon O. Eschenburg and Anne L. Eschenburg

My advisor,

Dr. Robert Kroeger

And my good friend,

Michael McPherson

Thank you for help and encouragement.

# Acknowledgements

I give my warmest thanks to the faculty and staff of the physics department of the University of Mississippi for their support and wisdom through the years. I especially thank the professors of the University of Mississippi's high-energy physics group; Dr. Jim Reidy, Dr. Don Summers, Dr. Lucien Cremaldi, and Dr. Robert Kroeger. Without your guidance and knowledge, this thesis would not have been possible. I also thank the many people of the SLD Collaboration who tolerated my many questions and were always willing to help me with my work. Lastly, I thank my fellow graduate students for their counsel, a source that was priceless.

# Abstract

The SLD experiment at the Stanford Linear Accelerator Center had a significant gap in its muon tracking coverage, provided by the Warm Iron Calorimeter. Supplemental planes of limited streamer tube chambers were added to improve the coverage in the vicinity of the gap at $0.65 < |\cos\theta| < 0.85$. A software effort to upgrade the tracking software for this system is detailed.

The commissioning of the forty-five degree chamber region of the SLAC SLD Warm Iron Calorimeter is presented. This task involved the completion of the forty-five degree chamber region geometry for the Warm Iron Calorimeter's fitter and swimmer and the changing of the way multiple scattering effects are treated in the fitter algorithm.

# Table of Contents

# List of Figures

**Chapter 1 - Introduction**

Albert Einstein's mass-energy relation is a well-known and frequently used equation in high-energy particle physics:

$$E = mc^2 \qquad \text{1-1}$$

Mass and energy are related. Lighter particles that have high kinetic energies when collided together can create heavier particles. A tool that observes this creation of new particles is called a detector. One such detector, the SLAC Large Detector (SLD) of the Stanford Linear Collider (SLC), was designed in the early 1980's. When electrons are collided against positrons at a sufficiently high momentum, $Z^0$ bosons are produced at rest. The purpose of SLD was to study the decay of the $Z^0$. [1]

SLD consisted of several major detector systems. Due to the matching energies of the electron and positron beams, the subsystems had a rotational and mirror symmetry. Shown in Figure 1.1, these are:

- Vertex Detector (VXD3) [2]
- Central Drift Chambers (CDC) [3]
- Cerenkov Ring Imaging Detector (CRID) [4]
- Liquid Argon Calorimeter (LAC) [5]
- Warm Iron Calorimeter (WIC) [6]

Unfortunately, the Stanford Linear Accelerator Center (SLAC) is located in a seismically active region. When earthquake factors were taken into account for the construction of SLD, a problem was discovered. It became apparent that shock absorbers needed to be placed between the endcap (EC) and barrel. This addition stabilized the detector in the event of a large earthquake.

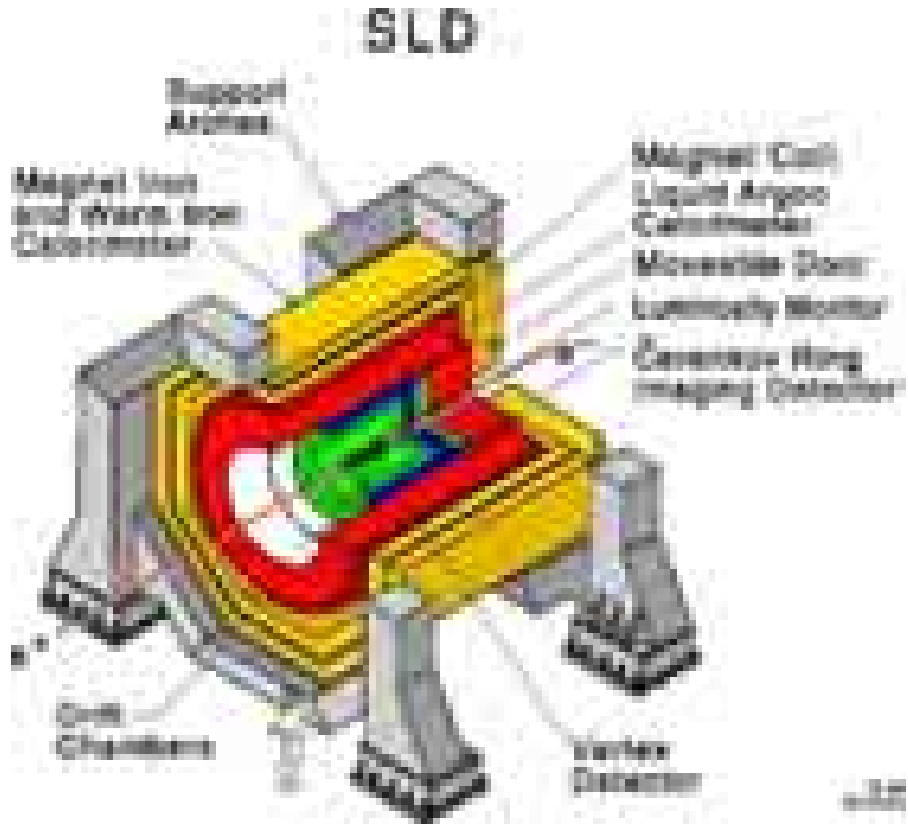

**Figure 1.1 - SLD cutaway that shows the various subsystems.**

After the detector was redesigned to meet earthquake safety standards, it was discovered that the SLD detector now had a significant gap in its muon tracking coverage.  SLD was like a soda can that is lying on its side. The barrel was the cylindrical part and the endcap areas were the two ends. The addition of the shock absorbers effectively pulled the ends outward. Muons were able to pass through this region and not be detected. This tracking efficiency gap was between $0.65 \leq |\cos\theta| \leq 0.85$, where $\theta$ is the angle between the beam line and the direction of the particle. This is shown graphically in Figure 1.2. The figure plots single muon identification efficiency versus $|\cos\theta|$ in Monte Carlo data.

The solution to improve the coverage consisted of adding supplemental Iarocci tubes around the gap region at forty-five degrees. These 45-degree detector planes were installed into any available space on the exterior of the WIC structure.

The completion of the software with these additions and its integration into the main analysis software of SLD was needed. After considering and fixing many of the problems inherent in the 45DC software, a notable improvement in muon tracking efficiency was achieved. The venture improved muon identification. This thesis describes the effort to commission these chambers.

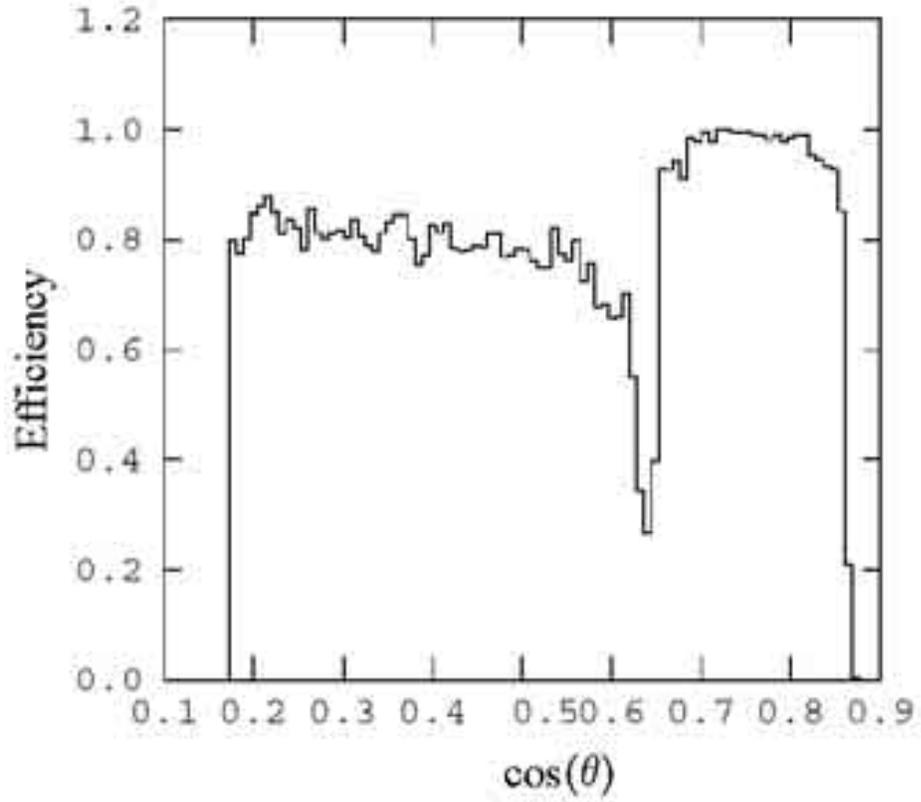

**Figure 1.2 – Monte Carlo generated muon tracking efficiency as a function of cos(θ). This figure shows the lack of coverage in the 45DC region.**

**Chapter 2 – SLAC, SLC & SLD**

**2.1 Introduction**

The SLAC Large Detector (SLD) was located at SLAC, a facility that adjoins the Stanford University campus and is situated thirty miles south of San Francisco, California. This 426-acre facility is also the location of other major experiments such as BABAR. [7] Completed in 1989, the Stanford Linear Collider (SLC) was built as an addition to the existing linear accelerator. At SLC's interaction point, SLD started taking data in the space formerly occupied by the Mark II detector. Using the existing electron/positron accelerator reduced SLD's footprint to SLAC.

Once the electrons and positrons are created in groups called bunches, they are then accelerated over a distance of two miles. The bunches form a line called a train and are bent by magnetic fields into separate beam lines and then steered into separate arcs. The beams are bent and focused by additional magnetic fields until they are brought to collision at the interaction point (IP). SLD was located symmetrically about this IP. An overview of SLC is shown in Figure 2.1.

To produce the $Z^0$ particles, the electron and the positron beams both have energies of 46.6 GeV. This does not combine to form the $Z^0$ mass of 91.187+/-0.007 GeV. SLC operated slightly above the resonance to optimize the polarization of the electron bunches. This polarization depended upon the electron trajectories and their energy. Running off resonance at high polarization helped to reduce the uncertainty of the measurement of certain quantities such as the left-right electroweak asymmetry parameter, $A_{lr}$.

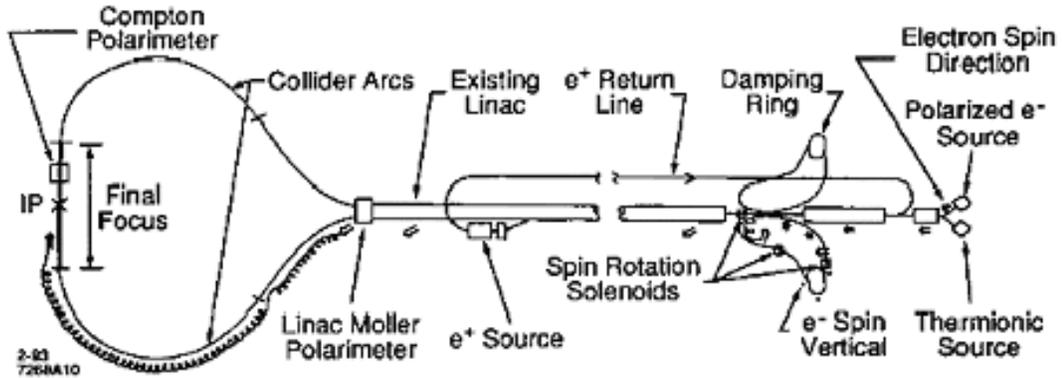

**Figure 2.1 - Overview of SLC.**

As shown in Figure 2.1, the SLC/SLD system allowed only one opportunity for interaction. After crossing the IP, the bunches were dumped rather than brought back. The benefit of this arrangement was that there is no need to replace the energy lost by synchrotron radiation, the energy radiated during the acceleration of a charged particle. Though the energy loss in the single passage in the SLC arcs was greater than the energy loss in one orbit in a storage ring, this energy loss was ignored because the electrons and positrons were only used once.

This reduces operating costs for running a linear accelerator. The cost of building and running linear colliders are linearly proportional to the energy of the particle being accelerated. The cost of building and running a storage ring environment are proportional to the square of the energy of their accelerated particles. As a result, linear colliders are a cheaper choice for studying the $Z^0$ from an energy consumption perspective. And as the center of mass energy of the accelerator increases, the choice of a linear collider becomes a better one.

Another benefit of a single-pass beam was the opportunity to focus the beam tightly. This allowed the vertex detector subsystem to be closer to the IP. Therefore, a better vertexing ability was achieved.

The acceleration of the positrons and electrons occurred within several regions of the SLC. The electrons were produced by a polarized electron source. At rates up to 120 Hz, two bunches, each consisting of $2\text{-}3 \times 10^{10}$ particles, merge into the beam line with an accelerating positron bunch after being stored in the cooling ring, an area that brings all the particles in the bunches to the same energy. [8] To produce positrons, one of the electron bunches, the scavenged bunch, was diverted from the main accelerator after it reached 30 GeV and then steered to collide with a rhenium-tungsten target. Showers of unpolarized electron-positron pairs were produced. These positrons were then collected, stored briefly in a damping ring, and then returned to the beam line.

In order for the electron/positron bunches to have uniform energies, SLC took advantage of the synchrotron radiation effect. When the particles were in the damping rings, the higher energy particles radiated faster than the ones with lower energy. The particles in the beam moved towards a homogeneous energy as time progressed.

Once the particles had a common energy, they entered the same two-mile long accelerator and were accelerated to 46.6 GeV. Positron and electron bunches alternated with each other on opposite phases of the machine cycle. At the end of this segment, the electron and positron bunches were separated by a uniform magnetic field. The polarized electrons were steered into the southern arc while the positrons were guided into the northern arc.

At the end of the arcs, the beams were focused to the dimensions of $2.6 \times 10^{-3}$ μm by $0.8 \times 10^{-3}$ μm by the time they reach the IP. Some of the collisions produced $Z^0$ particles at rest. Having such a narrow and stable interaction point allowed SLD to have very accurate vertex detection. The vertex detector was closer to the IP than is normally possible.

A unique quality of the electron beam in SLC was that it was partially polarized. Circularly polarized light was shined upon a strained lattice of Gallium Arsenide. This produced polarized electrons. [9] The electrons were accelerated with transverse spin and then steered so that their spin axis was lined up parallel to or anti-parallel to the beam axis when it entered the interaction point.

**2.2 SLD Subsystems**

**2.2.1 Vertex Detector (VXD3)**

When a collision occurs, many neutral as well as charged particles emanate from the IP. It is necessary to identify and measure their energy precisely. As shown in Figure 2.2, the first of these systems to study the event was the vertex detector. It was at the center of SLD. Three concentric tubes, with radii of 2.8 cm, 3.8 cm, and 4.8 cm each and 9 cm long, surrounded the interaction point. Shown in Figure 2.3, this cylinder used 96 charged coupled devices (CCD). These CCD's were like those used in digital cameras. Each CCD contained an array of 800 by 4000 silicon pixels at the surface of the vertex detector. The area of each pixel was 20 by 20 μm$^2$.

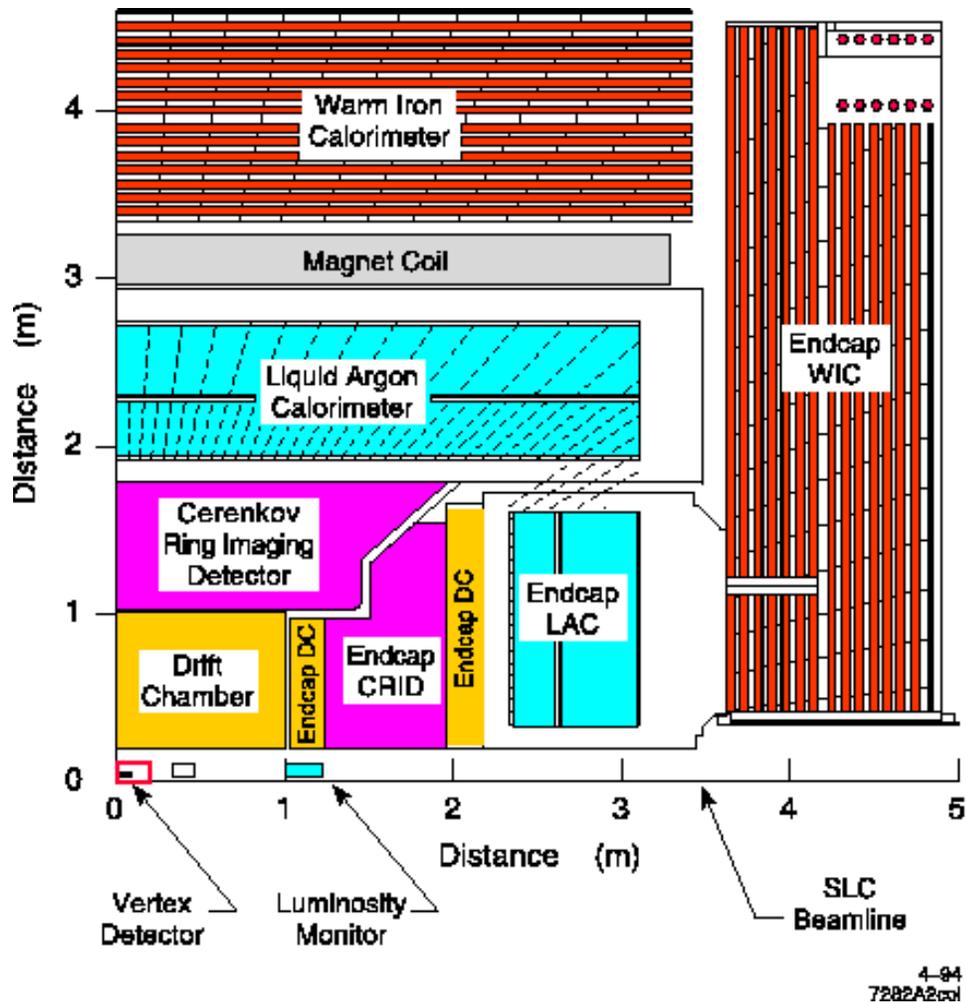

**Figure 2.2 – Cross section of one quadrant of SLD. The interaction point is shown at coordinates 0,0.**

When a charged particle passed through the pixels, ionization occurred in the silicon layers and then the separated charge is measured. This process takes about 50 ms. The vertex detector consisted of three nested barrels. Each segment of the barrel was 1 CCD wide and 2 CCD's long. This layering produced an overlapping of CCD's. The angular coverage of the VXD3 extended to $|\cos \theta| < 0.85$. The 1 σ single-track resolution of the VXD3 was ~350 μm.

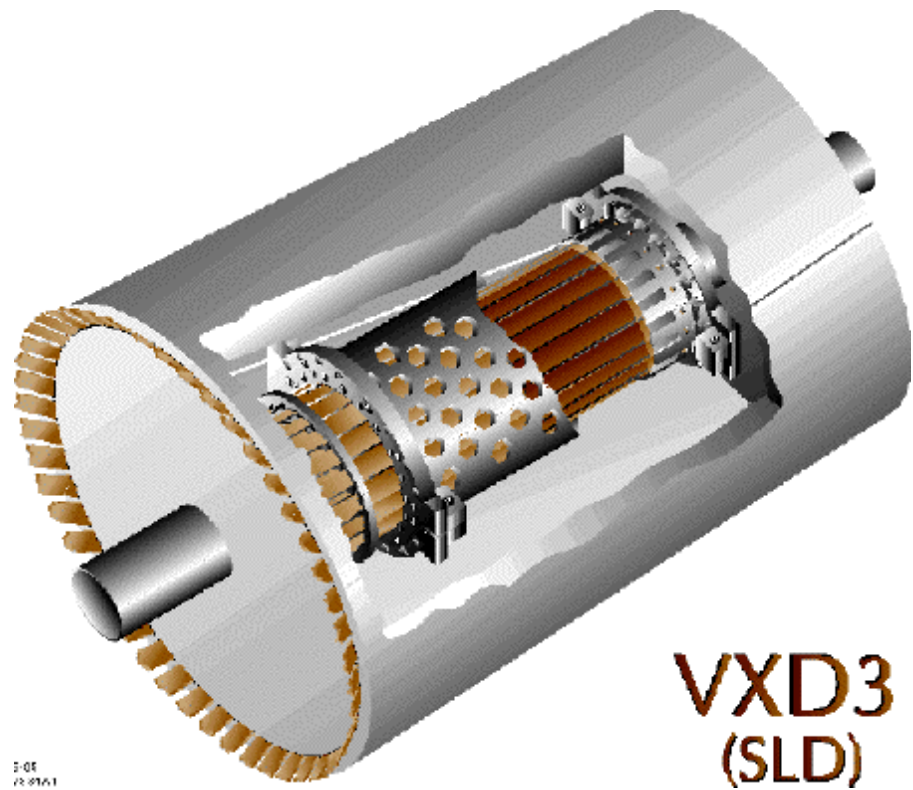

**Figure 2.3 – VXD3, the vertex detector of SLD.**

The vertex detector was cooled to –80º C with nitrogen gas. There were two reasons for this. The first was to reduce the noise caused by thermal excitations within the VXD3's silicon. And second was to maintain the CCD charge transfer efficiency that is adversely affected by radiation. The CCD passed its information passes along the data in a 'bucket brigade' fashion. The charge passed from one to another until it reached the end of the row. A higher temperature caused the charge stored in the CCD to be inefficiently passed and a loss of information to occur.

**2.2.2 – Central Drift Chamber (CDC) and Endcap Drift Chamber (EDC)**

The Central Drift Chamber (CDC) and the Endcap Drift Chamber (EDC) surrounded the vertex detector. The CDC was a cylinder that was two meters long, had an inner radius of 0.2 m and an outer radius of 1.0 m. Located inside a magnetic field, the CDC was used to determine the momentum of the charged particles.

Within the drift chambers a series of wires were strung. Some of the wires are high voltage wires. The remaining wires were connected to charge amplifiers. These remaining wires were also known as sense wires. The CDC and EDC were filled with a mixture of carbon dioxide, argon, isobutane, and trace amounts of water. The gas was ionized as the charged particle passes through the region. As the freed electron approaches the sense wires, showers of secondary electrons were produced in the gas. The sense wires produced a signal that was amplified, digitized, and then stored as data. This detector was in a 0.6 Tesla magnetic field, which was used for momentum analysis, and possessed a resolution of 100 μm. This allowed the CDC to resolve tracks that were within 1mm of each other.

## 2.2.3 - Cerenkov Ring Imaging Detector (CRID)

Once the particles leave the CDC or EDC, they passed through the Cerenkov Ring Imaging Detector (CRID). This detector identified the final state particle of the event. When charged particles pass through matter faster than light can move through that medium, light is produced. This process is called Cerenkov radiation. As time progresses in the CRID, a cone is formed. The angle at which the cone opens up, the opening angle, depended upon the velocity of the moving particle relative to the speed of light in that medium. This angle was:

$$\cos\theta = \frac{c}{vn} \qquad 2.1$$

where v is the velocity, c is the speed of light in a vacuum and n is the index of refraction in the medium. [10]

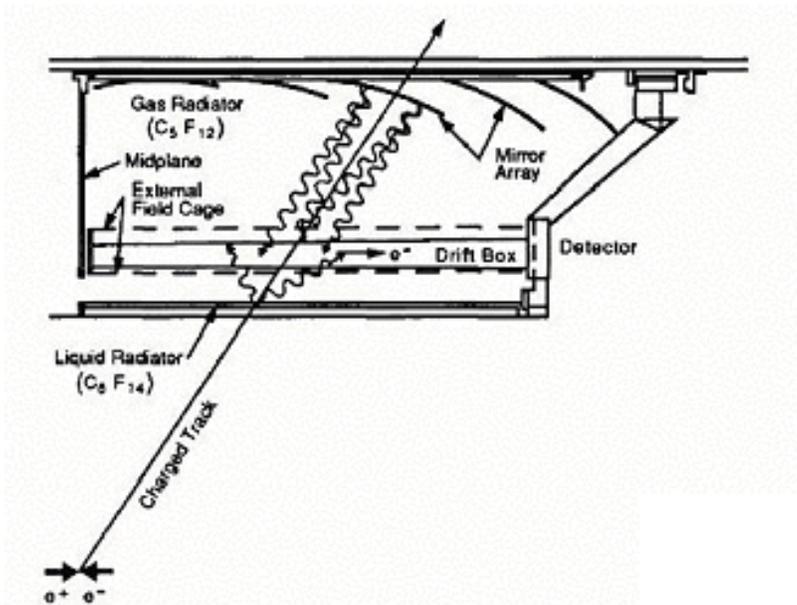

Figure 2.4 - CRID operation.

The momentum of the particles that traveled through the CRID varied widely. To cover this broad spectrum, the CRID used both gas ($C_5F_{12}$) and liquid ($C_6F_{14}$) Cerenkov radiators. Each radiator had a different index of refraction. Therefore each had a different range of the opening angle. As shown in Figure 2.4, a light cone emitted by Cerenkov radiation was seen as a ring in the drift box when a particle traveled through the thin layer of liquid. The particle then passed into the detector box and through a gaseous region. Parabolic mirrors mounted onto the drift box then focused the light cone created in the gas region into a ring in the drift box. The drift box was a multiwire proportional chamber with TMAE, a photocathode material, added to its drift gas to convert the photons. The resulting electrons drift to the sense wires, cascade, and are read out in a manner similar to the CDC. The two rings were then studied. The opening angles of the light cones were determined. The charged particle's velocity was then deduced. When this information was combined with the momentum data collected from the CDC, the mass and identity (i.e. type) of the particle was determined.

### 2.2.4 - Liquid Argon Calorimeter

Approximately one third of all particles created by the $Z^0$ decay were neutral. To measure their energy, they must be stopped in matter. The Liquid Argon Calorimeter (LAC) did this. It had the same barrel and end cap form like the CDC and the CRID. It covered 98% of the solid angle from the interaction point. It was made of lead tiles separated by plastic. This is shown in Figure 2.5. The space between the tiles was filled with liquid argon. When the neutral particle passed through this region, it interacted with the lead tiles and charged particles were produced.

The LAC was divided into four layers (EM1, EM2, HAD1, HAD2). The first two lead layers were thin (2mm) and designed to measure photons and electrons with good resolution. The remaining two tile layers were thicker (6mm) and are used to measure showers from hadrons, which produced deeper cascades. With these layers, 95% of the energy of a hadronic $Z^0$ decay can be contained within the LAC. Ionization proportional to the incident particle energy was deposited in the liquid argon. A high voltage was applied to every other lead plate layer. The resulting charge was then collected on the tiles that were uncharged. It was then measured and energy information was extracted.

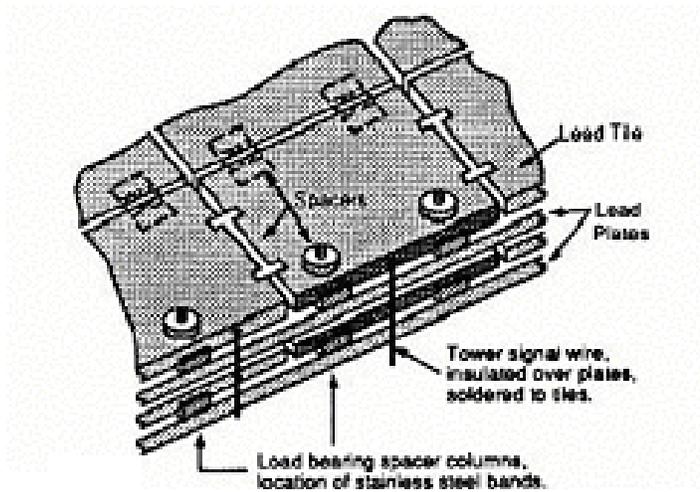

**Figure 2.5 - The cross section of the LAC.**

**2.2.5 The Warm Iron Calorimeter**

The exterior layer of SLD was the Warm Iron Calorimeter (WIC). As shown in Figure 2.6, it also had the same barrel and end cap form as the CRID and LAC. The barrel was comprised of Iarocci gas tubes layered between steel plates 5 cm thick. The primary function of the WIC was to track muons with a resolution of less than $1 \times 10^{-2}$ radians. The WIC was called a calorimeter because calorimetry was part of its intended purpose. It was later found that the data gathered was not useful for energy measurements. The WIC's primary purpose was then muon tracking.

Since this thesis deals with the commissioning of part of the WIC, this system will be discussed in greater detail in the next chapter.

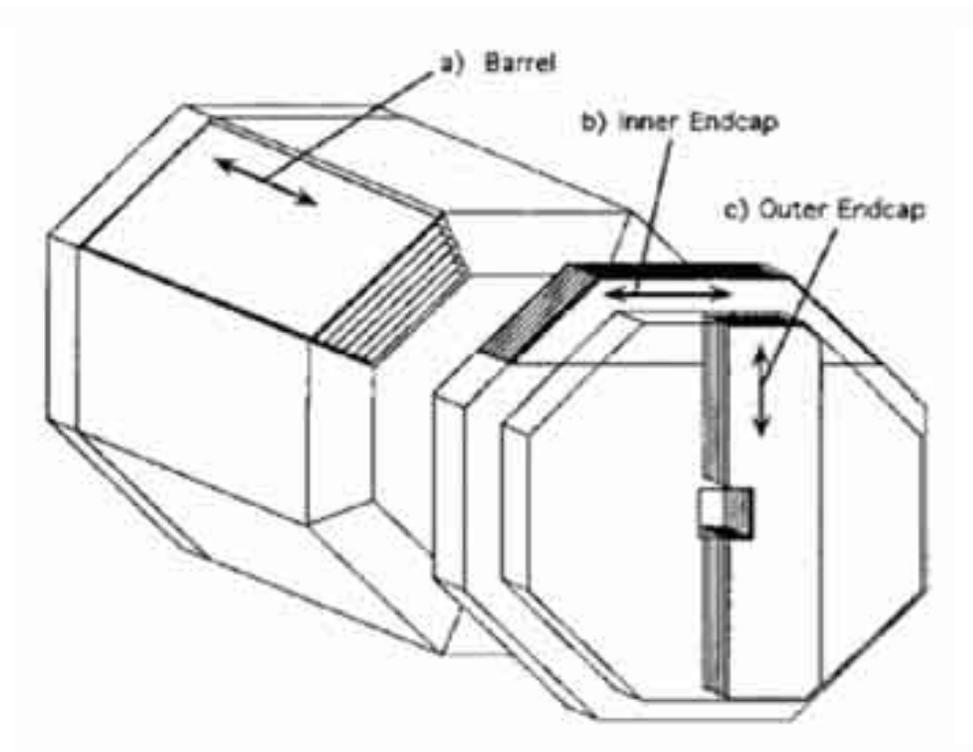

**Figure 2.6 The WIC barrel and endcap.**

**Chapter 3 – Detailed Description of the Warm Iron Calorimeter**

The Warm Iron Calorimeter (WIC) was the outermost layer of SLD. It was comprised of slabs of iron separated by planes of plastic streamer tubes. These tubes were also known as Iarocci tubes. [11] There were three subsystems to this detector subsystem. They were the WIC barrel, the end cap WIC and the forty-five degree chambers region (45DC).

The WIC served three purposes. The first was to track muons with the plastic streamer tubes. The iron served the second and third purposes. The iron was a path for magnetic flux return and stopped the remaining hadronic showers.

The total area that the limited streamer tubes cover was ~4500 $m^2$. This area included the tubes in the barrel as well as the end cap region. From the entire WIC, 80,000 strip and 8,000 tower channels were read. The process of reading the entire WIC took 0.8ms.

As shown in Figure 3.1, the Iarocci tubes each consisted of eight wires of high voltage that were suspended in a gas mixture at atmospheric pressure. When a particle passed through the cells, positive ions were produced and moved by the electric field to the inner surface of the tube. When the charge became sufficient, it was read by the electrodes mounted on the glassteel, a material made from glass fibers encased in epoxy. The passing particle could cause charge buildup on several electrodes. This was conducted by the graphite paint inside the tubes. They distributed the ion charge produced. The wires were run in limited streamer mode so the charge distribution was limited to a small length and the total charge was independent to the amount of the energy lost by the particle.

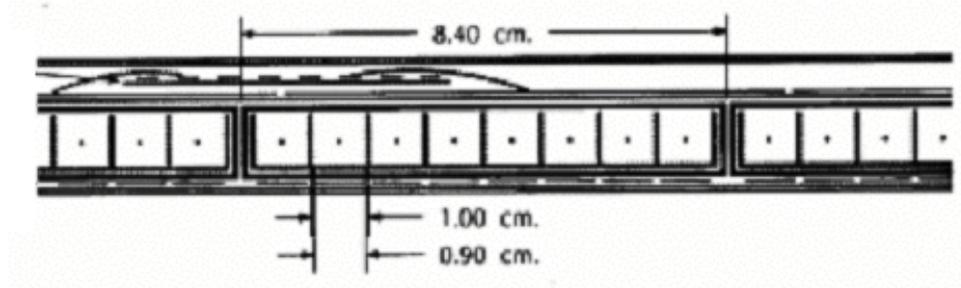

Figure 3.1 - Cross section of a set of Iarocci tubes.

The gas cell dimensions were 0.9 cm by 0.9 cm. The wire's diameter was 100 μm. The gas mixture was 88% carbon dioxide, 9.5% isobutane, and 2.5% argon. The voltage of the wires was 4750 V. The Iarocci tubes were placed in groups side by side to form the WIC chambers. Figure 3.2 illustrates their construction. Used for a platform for readout and structural stability, glass fibers encased in epoxy were attached to the tubes.

The barrel geometry consisted of eight octants. [12] The structure of an octant is shown in Figure 3.3. And, as shown in Figure 3.4, each octant was made of two coffins. These two coffins were staggered to eliminate regions of inefficiency between the octants. Each coffin had a height of 54 cm and was 315-360 cm wide and 680 cm long. The coffin was made of seven layers of 5 cm thick plates of steel that are separated by 3.2 cm gaps. Fitting in these gaps, the Iarocci tubes were run parallel to the beam. This accommodated the need for access to the tubes by the high voltage, gas, and instrumentation lines.

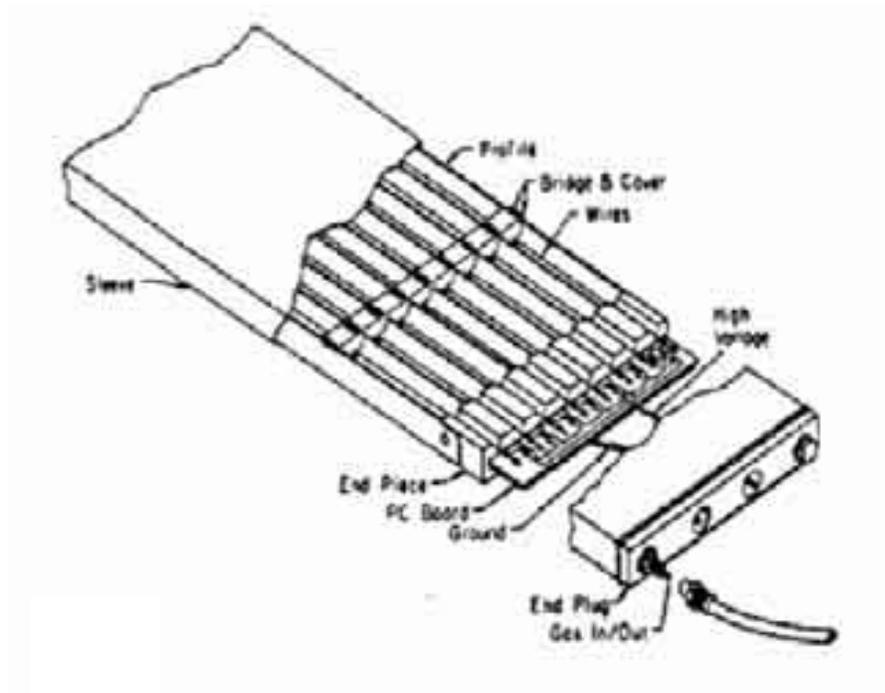

**Figure 3.2 – Iarocci tube construction for the EC and the 45DC and barrel.**

Each space in the steel layers of the WIC barrel contained one layer of Iarocci tubes. But double layers of these chambers were placed on the inner and outer layers of the outer coffin. Within this double layer is an outer layer of transverse strips perpendicular to the beam and the inner layer was built with longitudinal and transverse strips. The tubes were staggered by half a tube cell width from each other to eliminate geometric inefficiencies.

Both endcaps had six layers of chambers. But the outer endcap had two additional double layers of chambers on the inner and outer sides of the endcap. In total, the end cap had 18 layers: 6 horizontal and 6 vertical longitudinal layers, two vertical double width longitudinal layers, and four horizontal transverse layers.

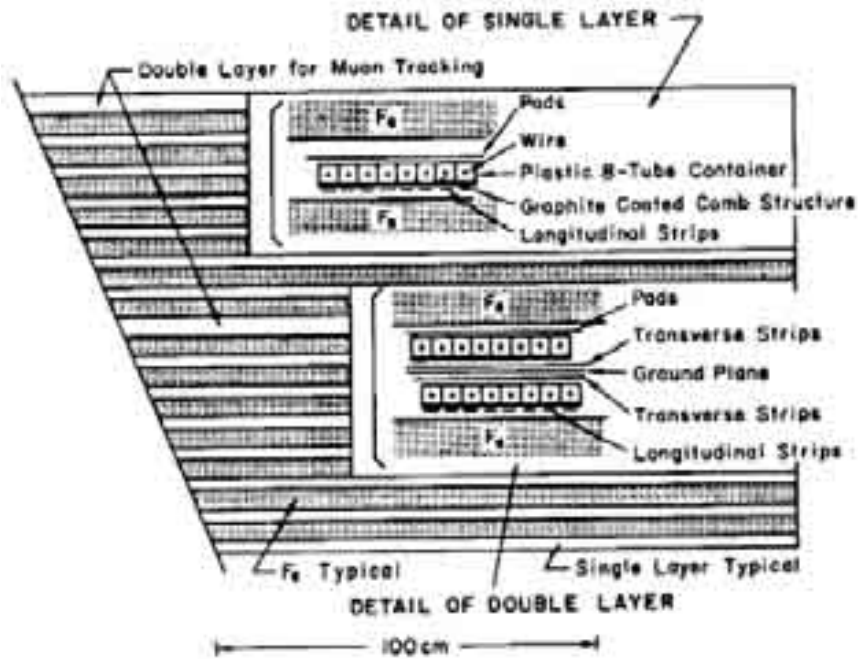

**Figure 3.3 – WIC barrel chamber construction[1].**

The barrel had fourteen longitudinal strip layers inside the coffin that were all parallel to the z-axis to measure the track momentum. It also had four layers of longitudinal strips that were arranged in two groups of two. These were used to measure the z position, the distance from the IP in the direction of the beam, of the muon.

---

[1] The 'pads' in this figure were intended for use in calorimetry but were never used. They have no role in the muon reconstruction effort described in this thesis.

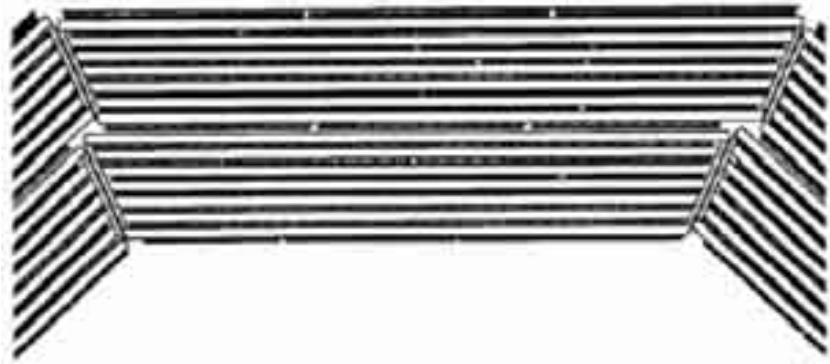

**Figure 3.4 – Sample of one octant of the WIC barrel.**

The third region of the WIC was the 45DC. The locations of the 45DC chambers are outlined in yellow in Figure 3.5. The 45DC chamber planes were positioned exterior to the steel support structure in the region of the seam between the barrel and end cap. Some were also positioned inside the legs that support the endcaps. Like the barrel, the 45DC was divided into octants. Unlike the barrel and the end cap, there was very little regularity in the placement of the steel. Though they do have forward backward symmetry, the locations of the chambers were also irregular in their placement. They were added post hoc taking advantage of every possible space. The total number of layers in this region was smaller than the number of layers in the other two regions. But in the 45DC, all of the chambers were double layered with Iarocci tubes having both transverse and longitudinal strips.

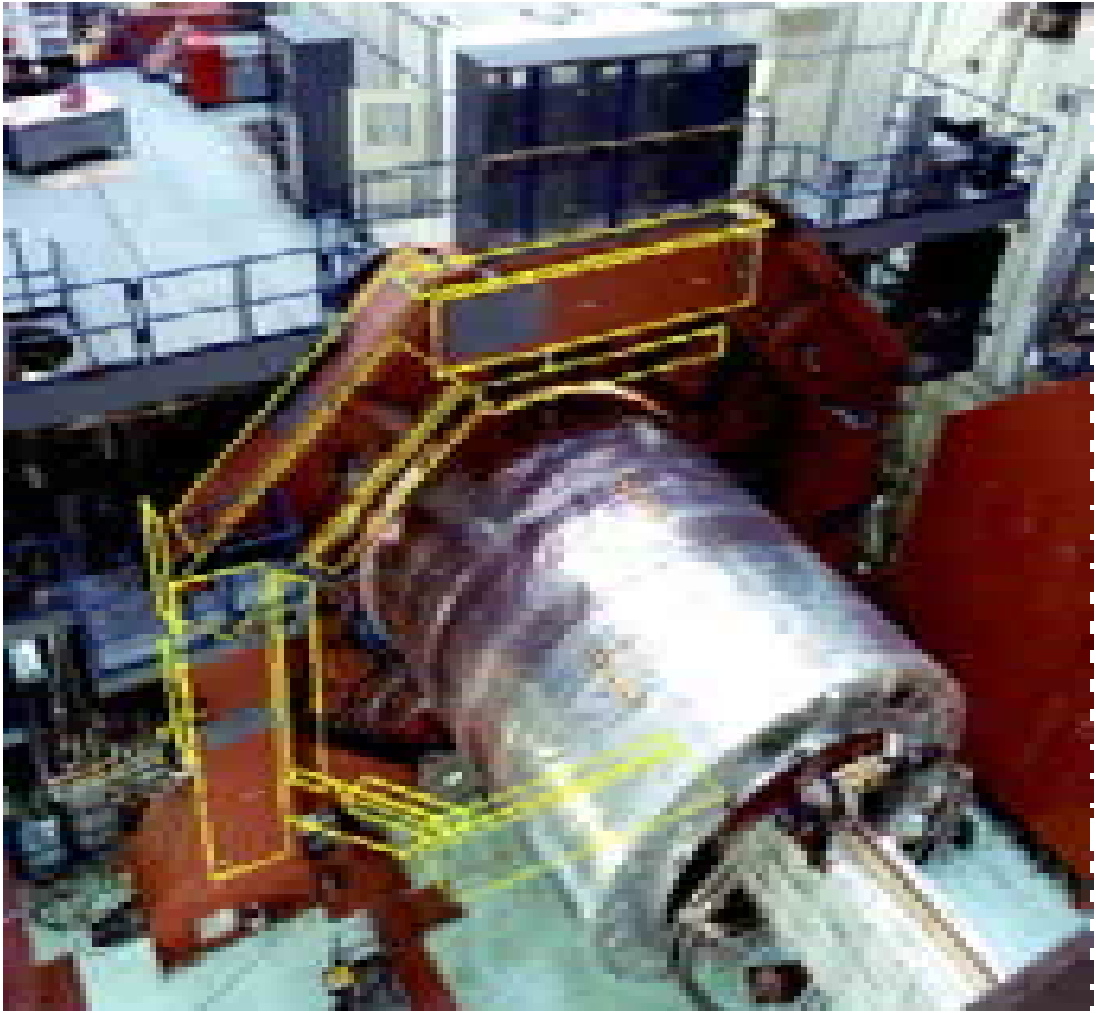

**Figure 3.5 – Outline of the 45DC. The silver cylinder is the LAC being inserted into the WIC.**

**Chapter 4 – Commissioning of the 45DC**

**4.1 – Introduction to the WIC Tracking Algorithms**

In every event, numerous particles passed through SLD. As the particles and their daughter particles moved through the detector, various subsystems recorded the progress. The resulting 'hits' were then digitized and stored. Afterward, these 'hits' were reconstructed into three-dimensional representations of the tracks. It was the WIC fitter that built the path of the particles through the Warm Iron Calorimeter.

When the event occurred, the data was stored in the forms of hit clusters. These clusters provided measurements along one spatial axis in the plane of the chamber layer through which the particle passed. These clusters were arranged into sequences called track elements. The WIC offered two readouts in two orthogonal directions of the particles track. The barrel and 45DC detectors provided strips that measured roughly the azimuthal and polar angle. The EC provided horizontal and vertical measurements. These data points provided the foundation for the reconstruction of the track through the WIC.

The first step in fitting the event with the reconstruction software was the use of the hit finder. Hit finding is the process of eliminating the hits that are unrelated to the track and selecting those that are. The hit finder formed track candidates out of all combinatorial possibilities that use no more than one hit in each layer and are consistent with multiple scattering criteria. The 'fitting' processes are the steps that study the remaining hits and then reconstructs the path of the track.

The hit finder selected particular hit clusters from various hits in a layer. At the beginning of the hit finding and fitting processes, the CDC passed along information concerning tracks from the CDC that can pass through the WIC. These were then extrapolated through the WIC. The WIC hits were then matched to these CDC tracks according to simple multiple scattering criteria. Any hits that were unrelated to the proposed tracks were eliminated. It then found the associated measurement axis from strip data for a track element. Uncertainties from the CDC tracking are added in quadrature with the uncertainties due to multiple scattering. And the hits used for a possible track must be within $4\sigma$ in two physical dimensions with respect to the extrapolated point of the track. And they are also constrained to use only consecutive hits that satisfy the requirement of being aligned to the scattering points within an error limit determined by the intrinsic error of each hit, the distance between hits along the scattering direction and the momentum. These track elements were then stored.

The remaining hits in the WIC were fitted into candidate tracks and those track candidates with large chi-squared values were discarded as described below. The remaining tracks were ranked according to the $\chi^2$ value, number, and variety of hits that they used. Those remaining tracks were then tested to see if they were muon tracks because they must be fully penetrating (they pass through the WIC completely) and identified as a charged track.

In the fitter step, a maximum likelihood fit was done (as described later in the chapter) to the four fit parameters. These parameters were:

- Two positions
- Two direction angles

An inverse momentum times charge may also be fitted if it is desired, but the momentum is ordinarily taken from the drift chamber tracking information.

For tracks in the barrel, the number of fit parameters is usually equal to four. If the fitter cannot resolve one or more of the fit parameters due to a lack of sufficient hits in the two orthogonal directions, one or more fit parameters must be excluded, and the number of fit parameters is less than four. The addition of 45-degree information in some regions of $\theta$ and $\phi$ made it possible for the fitter to find all four fit parameters where only two or three would be possible with the barrel and EC alone. (See Figure 6.5).

Simultaneously satisfying the equations:

$$\frac{\partial \chi^2}{\partial P_i} = 0 \qquad 4.1$$

for each parameter $P_i$ yields the minimum values for the $\chi^2$. These simultaneous equations were solved via a linearized least square fitting procedure.

The chi-squared function of the proposed track was determined by propagating the track through the WIC. This was a stepwise integration of the equations of motion taking into account energy loss, deflection by the magnetic field, and the incremental changes to the covariance matrix using the local values of the radiation length as described below. The expressions for values of the position $\vec{r}$ and direction $\hat{t}$ after the particle has traveled a distance s is [14]:

$$\hat{t}(s) = \frac{d\vec{r}(s)}{ds} = \hat{t}(0) + q \int_0^s ds' \hat{t}(s') \times \frac{\vec{B}(s')}{p(s')} \qquad 4.2$$

$$\vec{r}(s) = \vec{r}(0) + s\hat{t}(0) + q \int_0^s ds'(s-s')\hat{t}(s') \times \frac{\vec{B}(s')}{p(s')} \qquad 4.3$$

And the expression for the momentum on s that takes into account the energy loss due to continuous electromagnetic processes is:

$$p(s) = p(0) - \int_0^s ds' \kappa(s') \qquad 4.4$$

It is this numerical piecewise integration of these equations that is called 'swimming'. The predicted values of position and momentum from the above equations are determined and then used in the calculation of the chi-square function. These equations took into account the geometry of the WIC, the local magnetic field, the radiation length, and the loss of energy throughout the regions. This is why the geometry of the region and its composition (air, steel, or chamber) has to be known.

A chi-squared value is formed:

$$\chi^2 = \sum_{i,j=1}^{N} \left(y_i^* - y_i(\{P\})\right) W_{ij} \left(y_j^* - y_j(\{P\})\right) \qquad 4.5$$

Where:

- $y_i^*$ is the $i^{th}$ of the N measured coordinates that constitute the track,

- $y_i(\{P\})$ is the fitted function of the parameters corresponding to $y_i^*$

- The parameters set $\{P\}$ consists of positions and slopes of the initial trajectory projected along two orthogonal directions on a specified reference.

- $W_{ij}$ is the weight matrix.

To find the minimum $\chi^2$ as a function of the fit parameters, the equation used to calculate the chi-squared is expanded in Taylor series about its minimum. The linearized problem is then solved by iterative linear algebra methods. This expansion is [16]:

$$\chi^2(P+\Delta P) = \chi^2(P) + \Delta P_\alpha \frac{\partial \chi^2(P)}{\partial P_\alpha} + \frac{1}{2} \Delta P_\alpha \Delta P_\beta \frac{\partial^2 \chi^2(P)}{\partial P_\alpha \partial P_\beta} + \ldots \qquad 4.6$$

The minimum of the $\chi^2$ under this approximation would be found at the point where the parameters set $\{P\}$ is given by:

$$P^{stat}_\alpha \approx P_\alpha + \Delta P_\alpha = P_\alpha - \left(A_{\alpha\beta}\right)^{-1} B_\beta \qquad 4.7$$

$$\text{where } B_\beta = \frac{\partial \chi^2(P)}{\partial P_\beta} = 2\frac{\partial x_i}{\partial P_\beta} W_{ij}(x_j - x_j^*) \qquad 4.8$$

$$\text{and } A_{\alpha\beta} = \frac{\partial^2 \chi^2(P)}{\partial P_\alpha \partial P_\beta} = 2\frac{\partial^2 x_i}{\partial P_\alpha \partial P_\beta} W_{ij}(x_j - x_j^*) + 2\frac{\partial x_i}{\partial P_\alpha} W_{ij} \frac{\partial x_j}{\partial P_\beta} \qquad 4.9$$

The fitter then goes through the process of adjusting the direction and fit parameters to these new values calculating the $\chi^2$ at the new parameters and its derivatives, and iterating.

Due to multiple scattering, the weight matrix has non-diagonal elements. The relationship between the measurement residues at the ith and jth detector plane due to these multiple scattering deviations are represented in the elements of the inverse weight matrix, $W_{ij}^{-1}$:

$$W_{ij}^{-1} = \left\langle \left(y_i^* - \left\langle y_i^* \right\rangle\right)\left(y_j^* - \left\langle y_j^* \right\rangle\right) \right\rangle - \left\langle y_i^* - \left\langle y_i^* \right\rangle \right\rangle\left\langle y_j^* - \left\langle y_j^* \right\rangle \right\rangle$$

$$= \delta_{ij}\sigma_i^2 + (14\text{MeV}^2)\hat{m}_i \cdot \hat{m}_j \times \int_0^{\min(s_i,s_j)} ds' \frac{(s_i - s')(s_j - s')}{\left[\hat{t}(s_i) \cdot \hat{n}_i \hat{t}(s_j) \cdot \hat{n}_j X_{rad}(s')p(s')^2\right]} \quad 4.10$$

Where:

- $\sigma_i$ is the rms position resolution error assigned to the $i^{th}$ and measurement.
- $s_i$, $s_j$ are the path lengths to the points where the track crossed the $i^{th}$ and $j^{th}$ detector planes.
- $\hat{t}$ is the direction vector.
- $\hat{m}_i, \hat{m}_j$ are unit vectors along the $i^{th}$ and $j^{th}$ measurement axes.
- $\hat{n}_i, \hat{n}_j$ are unit vectors normal to the $i^{th}$ and $j^{th}$ detector planes.
- $p(s')$ is the momentum times c.
- $X_{rad}(s')$ is the radiation length at path length $s'$ along the trajectory.

The integral is taken during the swimming process over all the material in front of both detector planes The fitter was a core feature of the WIC software. To do its work, the WIC fitter needed a GEANT-like geometry. The positions, dimensions, and substance of every volume must be correct. These volumes were represented in the fitter by inequalities that represent boundary planes. The swimmer of the fitter integrates the path of the track according to the dimensions and material of the volumes in the detector. The track loses energy according to the material of the space.

To speed up the process of locating which volume contains the hit, 'pointers' are used in the volume search algorithm. If the point being studied fails an inequality test with a volume face, the pointer provides the next appropriate volume to be searched. The sides must be appropriately chosen so every space in the WIC region can be paired with its correct volume quickly.

**4.2 Necessary Upgrades**

The reconstruction software for the Warm Iron Calorimeter sub-system is voluminous; ten FTE years went into developing it. During this large effort very little thought was devoted to the eventual inclusion of the 45-degree chambers in the overall code scheme.

The barrel and end cap portions of the WIC were treated in the hit finding, the muon pair filter and WIC fitter as if they had been two separate muon spectrometers. The coordination of the two regions had, in some degree, been deferred to a subsequent integration with the 45-degree chambers.

Much of the WIC code was developed assuming certain simplifying features of the geometry, which the layout of the 45-degree chambers violated badly. The WIC barrel and endcap geometries were simple and regular, planes of chambers between plates of steel. All the physical volumes were thus a series of parallel, six-sided boxes in contact with their neighbors and arranged into an eight-fold symmetry. Their dimensions and locations could be easily reconstructed with a radial index, an octant number, and a transverse index. The 45DC region violated all of these tidy assumptions forcing drastic revision of the hit finding and WIC fitter code.

When compared to the rest of the WIC, the 45DC is very irregular. The 45DC contains approximately one hundred chamber layers and many steel beams that are oriented in numerous different directions. Most of the chambers and steel segments do not align with each other. Lastly, these volumes in the 45DC region have as few as four sides and as many as ten sides and obey no octagonal symmetry.

For these reasons, code for the WIC muon tracking had to be updated in many aspects to include the 45DC. Upgrades to the muon pair filter, the GEANT geometry, the hit finder, the fitter/swimmer geometry, and the fitter algorithm were all implemented.

Only three of these major tasks are addressed in this thesis in detail. The first was the creation of an accurate geometry consisting of a space-filling set of convex volumes for the fitter/swimmer in the 45DC region. The second was the selection of proper order of the volume faces and appropriate 'pointers' for the volumes. The last task involved rewriting the fitter code to accept results from the 45DC and a new value of the 'obliquity' factor, which adjusts the multiple scattering correlations between measurements in different chambers due to their spatial orientations.

**4.2.1 – Numbering Convention Incompatibilities**

In the barrel and end cap, geometrical information was encoded into the subsystem number and layer number. In the 45-degree chambers these numbers were generally meaningless or conflicted in definition with their use in the barrel and end cap. Improvisations had to be made in many places to incorporate the 45-degree chamber hits into an overall scheme to which they were not at all well suited. All of the 100 chamber planes were treated as special cases. The 45-degree chamber information in the pattern recognition step was added by creating a new direction type for each 45 degree-chamber layer. This was the simplest work around to the problem that 45-degree chamber layer numbers would have otherwise conflicted with barrel and endcap layer numbers for strips in the same direction.

**4.2.2 – WIC Fitter Geometry**

The greatest difficulty associated with including the 45-degree chambers in the overall program of track fitting had to do with the fact that the WIC track fitter has its own geometrical description of the WIC geometry independent of the GEANT geometry. The fitter's locator routines assume that the detector geometry is a series of tightly fitting convex volumes, which have eightfold symmetry by octant. The 45-degree chambers violate these conditions, forcing drastic revisions to the fitter geometry and associated codes, which are described below.

The WIC track fitter originally required about three man-years of development, and the retrofitting necessary to accommodate the 45-degree chambers has proved to be a project similar in scope. WIC track fitter has its own geometrical description of the WIC geometry independent of the GEANT geometry. It also has an independent track swimming algorithm that depends on the one in GEANT only to the degree that it uses the GEANT description of the magnetic field.

### 4.2.3 – WIC Fitter Adjustments and Incorporation

The WIC fitter algorithm was reworked in one additional way beyond the geometry changes. Changes were made to the obliquity factor in the $\chi^2$ equation. This factor helps account for the differences in orientation of various chambers (normal directions and measurement axes) when calculating the effects of multiple scattering on the weight matrix. This work is described in the next chapter.

### 4.2.4 – Dead and Noisy Strips, SLC Background

In addition to the lack of attention the forty-five degree chambers were receiving from the software side, these chambers suffered the same treatment in terms of their care and maintenance in some degree. Over the years, many channels became 'hot', producing incessantly spurious signals (Figure 4.1). These were easily confused with the high backgrounds from SLC muons. The WIC dead and noisy channel management code had been disabled because the barrel and endcap WIC were so robust and well maintained that the experts had simply decommissioned the dead channel management with virtually no impact on the purity or efficiency of the sample. But for the 45DC data, since there was only a single plane in each type, this was not the case.

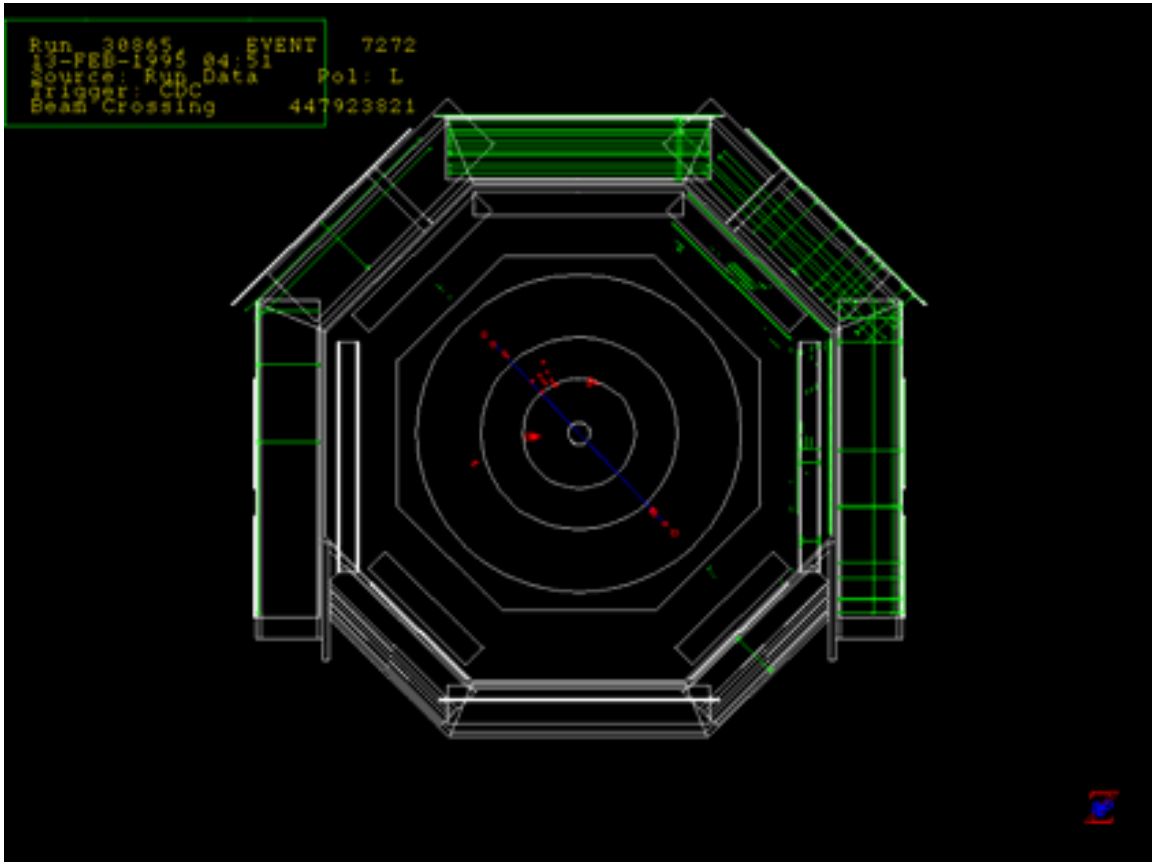

**Figure 4.1 – Event display showing hot strips, SLC Backgrounds.**

### 4.2.5 - Challenges

The four major challenges concerning this commissioning of the 45DC geometry and its usage by the WIC fitter were:

1. A significant amount of time had passed since the original WIC software commissioning. Both documentation and the original experts were hard to locate.
2. The 45DC was a large and geometrically complex system. Its geometry was highly incompatible with the existing WIC code as described above.
3. The WIC code was very extensive, having required 10 FTE years to create. The retooling to include the 45DC was a very major undertaking.

The physical dimensions and positions of the 45DC were determined after a number of steps. The original experts had moved to other projects and were generally not available for consultation. No blueprints or other authoritative documentation existed for the 45DC. The various pieces of documentation that did exist conflicted with each other. Most of the original GEANT description of the chamber was based on tape measure readings. With conflicting sources, the geometry was also checked with the hits created by Monte Carlo generated muons.

**4.3 – WIC Fitter Geometry Work**

Robin Verdier of M.I.T., the original architect of the WIC fitter, agreed to study the problem of a retrofit for the fitter code. The geometrical incompatibilities were so severe that Robin gave serious consideration to recommending that the existing fitter be thrown away and to start creating a new one from scratch. He eventually concluded that the best approach was to try to create a specification of the 45-degree chamber geometry to be incorporated into the existing fitter geometry along with new locator routines utilizing this special geometry and code.

The fitter was designed so that its locator routines would execute much faster than the comparable GEANT locator routines. The method associates a space point with the correct volume by determining on which side of a series of planes the point falls.

**4.3.1 – Volume Requirements**

The fitter geometry for the arch/leg and 45-degree chamber region had to be entirely enclosed in a headstone shaped region (to join the "coffins" in the barrel region), which is the smallest convex volume that contains all the steel and chambers. The volumes of the 45DC consisted of an exhaustive data file specifying the normal direction for each plane bounding each volume and its impact parameter to the interaction point.

The first requirement was that every cubic millimeter within the "coffin" had to be classified as steel, detector, or air. Once the open regions were determined, new volumes of air were created. Volumes were added and then the region was scanned again to find the remaining open regions. After all of the unclaimed space was occupied, conflicts between volumes were resolved. Locations that were claimed by two or more volumes had to be rebuilt. Each of the volume's corners, as well as many other points on the surface of the volumes, was checked to see if other volumes claimed them. Because of the uniqueness of the geometry, every region required its own set of solutions to the volumes. Our group took over and completed vestigial versions of the geometry (and the new locator routines), which were based on inaccurate drawings.

All the steel in octant 7 arch and leg structure had to be rebuilt a couple of times as better information about the geometry was obtained. This was very major surgery as the particular joint between octant 7 arch and legs includes an awkward intersection where chambers inside the end cap door legs (oriented at 48 degrees to the beam axis in the horizontal plane) meet vertical and 45-degree diagonally oriented steel. This is a region with many of irregularly shaped volumes some seven sided. Figure 4.2 illustrates this complexity.

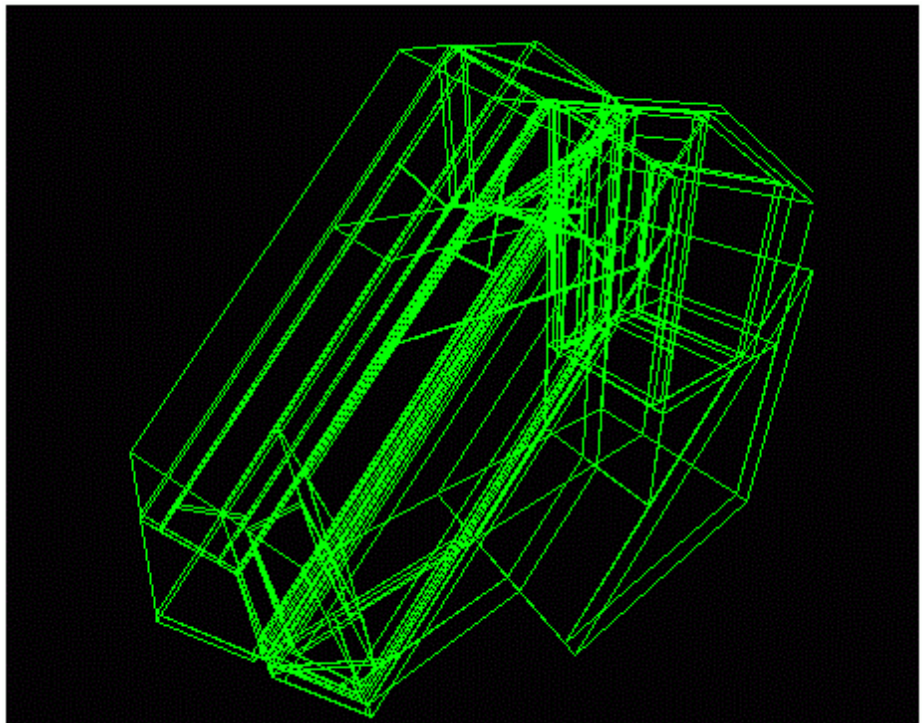
**Figure 4.2 – Octant seven region of the 45DC.**

### 4.3.2 – Pointer Requirements

With the fitter geometry completed the work on defining the geometry's pointers began. Having a complete set of planes is insufficient for the WIC fitter's requirements. Pointers assist in the fitter algorithm's determination of where the track is going. Not only must the pointers must be carefully chosen, the order of the planes for each volume must also be selected well. The specification of the pointers for the 45-degree geometry posed a difficult "traveling salesman"-like problem because of the irregular shapes of the volumes. There can be up to ten sides, and in some cases, no two sides are parallel or no two orthogonal. Though a given face of a volume may meet as many as 14 faces of other volumes, the software is limited to using one pointer for each face. This presented a challenge. It was problematic to insure that there are no regions unreachable by the search, and no regions that result in infinite loops. A point and click function to survey all the faces opposing any specified volume, a utility that is described later, was invaluable in avoiding potential trouble spots in the 1200 pointers assigned.

**4.3.3 – Volume Creation Tools**

The main visualization tool was an X-windows program that drew the volumes. It was constantly being rewritten to accommodate new needs. Routines were created to display the 45DC WIC fitter geometry by unpacking the direction cosine of the normal vector with respect to the beam line and impact parameter for each face of a volume, finding the vertices and drawing the faces. One of the added features that was rather helpful was one that visualized the volumes in three dimensions. Figure 4.4 shows the main visualization window in 3D mode. The viewer to see the image wore glasses of red and blue lenses. In addition, other analysis routines were created. Some sought cases in which a vertex fell inside another volume in order to locate conflicts between volumes, and an interactive editor function was built to alter the volumes on-line in the display program.

Initially, the layout of the 45DC planes in the vestigial fitter geometry that was inherited did not seem to match the event display very well. To investigate, supplemental code was built into the routines that construct the GEANT description of the 45-degree region and the arch support structure. Whenever GEANT routines created a trapezoid or box, this code calculated the direction cosines and impact parameters for comparison to the fitter geometry. There proved to be very extensive differences between the two descriptions; giving rise to a long process of adjudicating between the two geometries, rebuilding and re-rebuilding the geometry.

Initially, voids, small gaps, and conflicts were not readily discerned in the wire frame drawings. For these reasons, and in order to resolve a large class of conflicts in which two volumes cross but neither one has a vertex inside the other volume, a set of new diagnostic tools was developed.

Raster programs were designed to step through every point in the headstone region that measured in the hundreds of cubic meters millimeter by millimeter. A sample display of the raster data is shown in Figure 4.3. As the resolution of the scans improved, the time required to complete a scan increases in proportion to the cube of the inverse scale of the resolution. On only one occasion was a high-resolution scan of the entire region performed. It took twelve DEC stations [17] five full days to complete.

Points assigned to more than one volume were flagged as conflicts; those unassigned to any volume were histogrammed against each of the direction cosine axes so the outlines of missing volumes were identified by direction cosine and approximate impact parameter. Since each impact parameter is specified as an integer number of millimeters, small voids and conflicts of less than millimeter dimension must be tolerated wherever two volumes from neighboring octants meet at a point not on a 45-degree diagonal. The WIC fitter software doesn't tolerate gaps larger than a millimeter in size. A raster scan of the full 45-degree region would take months at that scale so fast and highly automated routines to survey the faces of a given volume for conflicts or adjacent voids without the long turnaround time were developed.

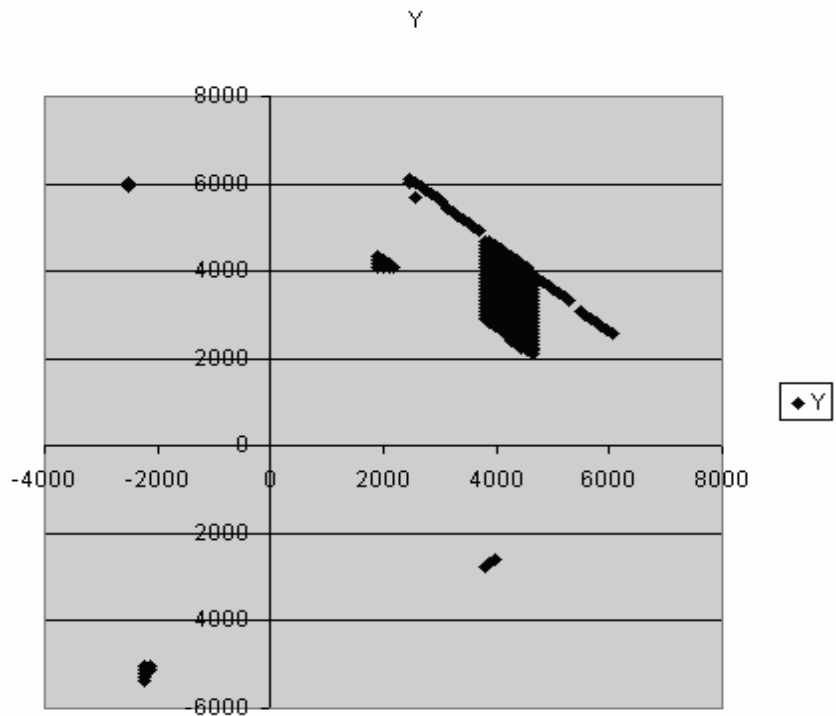

**Figure 4.3 – Output collected from raster scan that shows unclaimed space.**

A point-and-click function was developed to survey the neighbors of faces interactively to check for conflicts and neighboring voids. The routine rotates the view to the necessary local coordinates to display a selected face at normal incidence, draws this face and presents a crosshair. Shown in Figure 4.5, the user clicks a given point of the face and the routine returns the number of the neighboring volume. Some faces bordered as many as 14 neighbors.

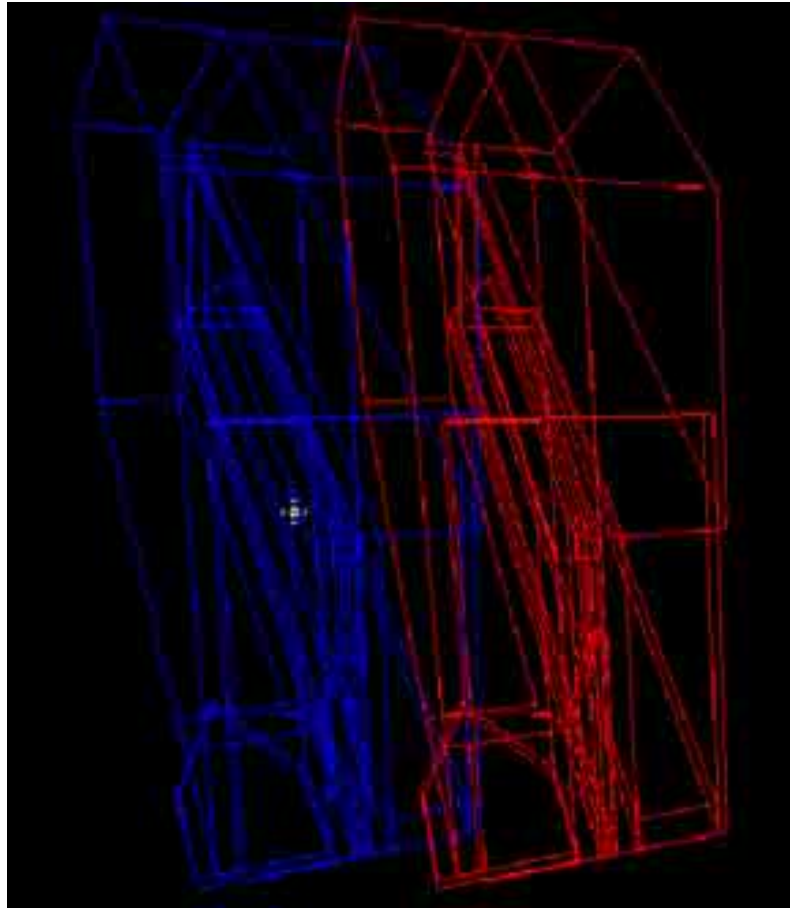

**Figure 4.4 – Visualization tool being used in 3D mode.**

The point-and-click function was useful but slow so it was automated via a triangulation scheme. A sample of this process is shown in Figure 4.6. This program applied several rules to every face of every volume. Each face is divided into triangles using the vertices of the faces as the vertices of the newly created triangles. These points are then moved 1 mm away from the face along the normal direction of the face being studied. These new points are then checked to see which volume occupies that space.

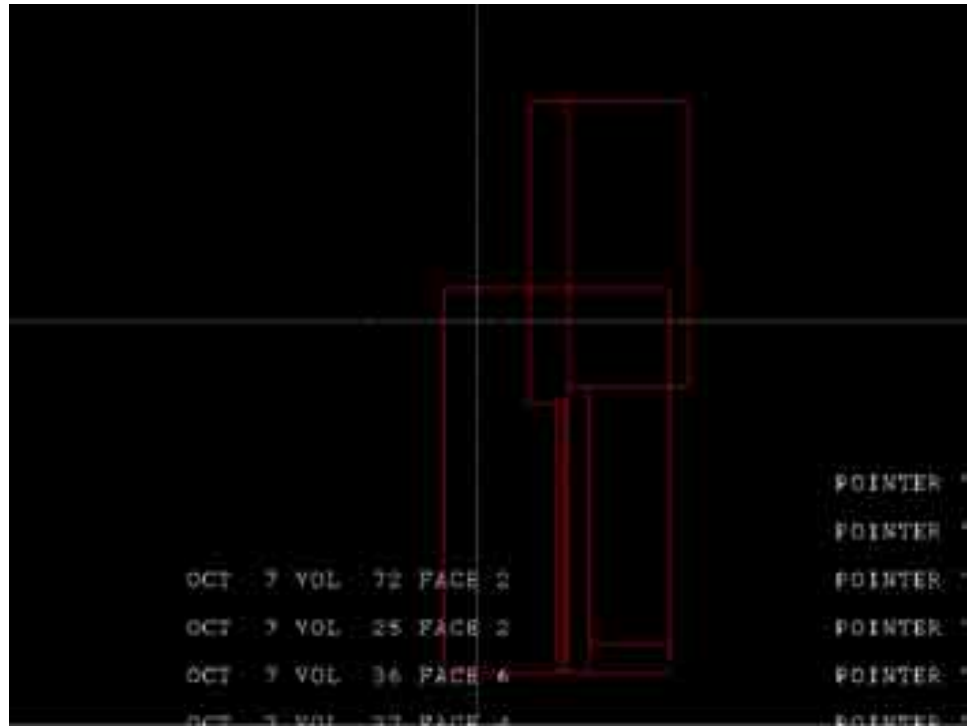

**Figure 4.5 – Visualization tool being used in 'Point and Click' mode.**

If the same volume claims all three points, it is no longer investigated further and the program moves to the next triangle. If no sides match, then that triangle is broken into two smaller triangles, breaking the longest side is broken at its midpoint. But if two vertices match, the triangle is broken along the intersecting plane that is between the two matching points and the third. And this process is repeated. Points that were never resolved were stored in a file for later study. This procedure was useful to list neighboring volumes and resolve conflicting regions within a short time. The whole 45-degree fitter geometry could be surveyed in ten CPU minutes on a millimeter scale for most types of conflicts and voids, as compared with months for the raster scans. The final version of the 45DC geometry is shown in Figure 4.7.

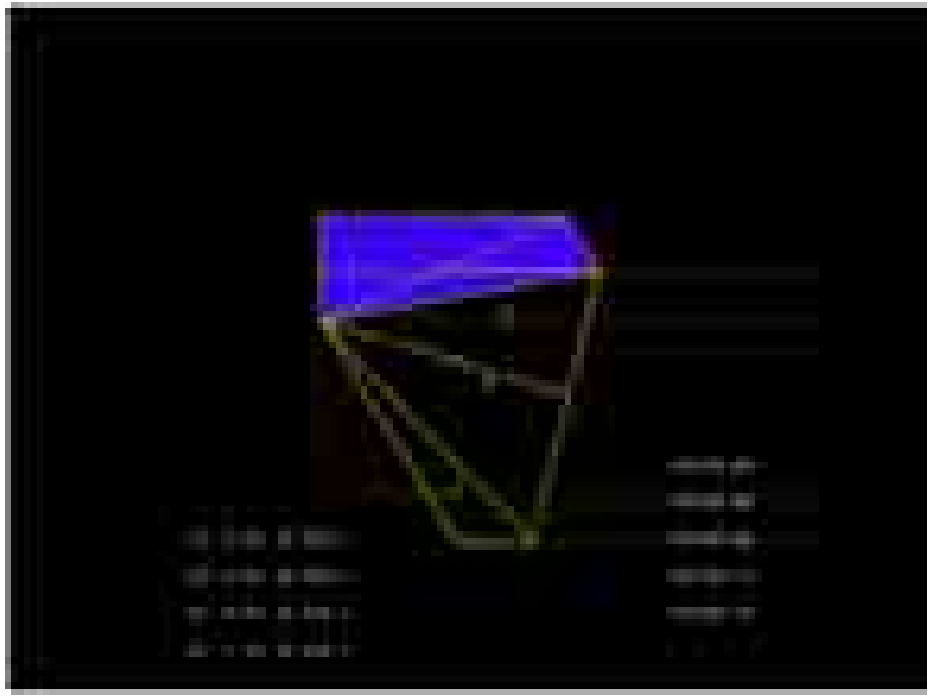

**Figure 4.6 – Triangulation scheme in progress. This picture shows a repeated construction of triangles and determination of opposing volumes.**

### 4.3.4 – Pointer Specification Tools

In the determination of the pointers of the volumes in the 45DC, the triangulation function was of the most use. From this program, opposing volumes were quickly determined. It was important to know all volumes opposing the side being studied so the appropriate pointer is chosen.

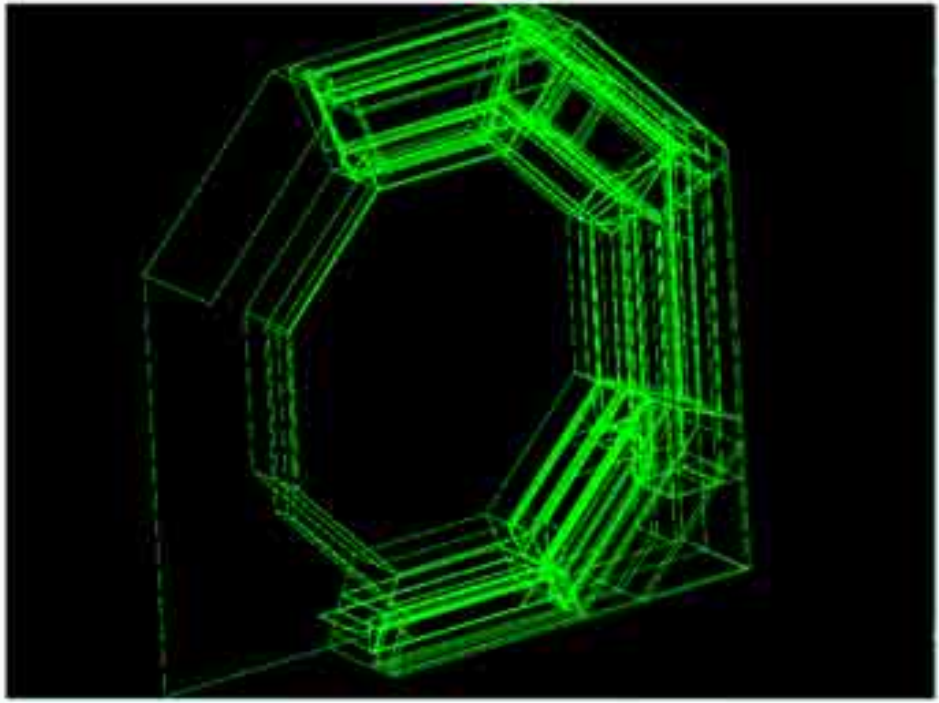

**Figure 4.7 – Final 45DC geometry for one of the towers of the 45DC. The volumes in the left part of the picture do not show because of the software's use of left-right forward-backward symmetry from the other octants.**

# Chapter 5 - WIC Fitter

## 5.1 - Description and Details of the Changes to the Fitter Algorithm

The WIC fitter provided increased purity over the pattern recognition stage based on fitter's careful treatment of correlations among successive wire hit residuals due to multiple scattering effects.

The expression for the weight matrix:

$$W_{ij}^{-1} = \langle (y_i^* - \langle y_i^* \rangle)(y_j^* - \langle y_j^* \rangle) \rangle - \langle y_i^* - \langle y_i^* \rangle \rangle \langle y_j^* - \langle y_j^* \rangle \rangle$$
$$= \delta_{ij}\sigma_i^2 + (14\text{MeV}^2)\hat{m}_i \cdot \hat{m}_j \times \int_0^{\min(s_i,s_j)} ds' \frac{(s_i - s')(s_j - s')}{\left[\hat{t}(s_i) \cdot \hat{n}_i \hat{t}(s_j) \cdot \hat{n}_j X_{rad}(s')p(s')^2\right]} \quad 5.1$$

has an expression within it that is known as an 'obliquity factor':

$$\frac{\hat{m}_i \cdot \hat{m}_j}{(\hat{t} \cdot \hat{n}_i)(\hat{t} \cdot \hat{n}_j)} \quad 5.2$$

This factor is appropriate for fixed target geometry with all the detector planes aligned along a common normal direction and with the track trajectory nearly normal to the detector planes in all cases. An attempt was made to improve the cooperation between the barrel end cap and 45-degree sections of the WIC by reconstituting the fitting algorithm itself.

In the barrel and end cap joint the assumption that a track will have a nearly normal incident angle with respect to the chamber plane's surface is not applicable.

Furthermore, the barrel, end cap, and 45DC do not have a common normal direction. Therefore, some pairs of hits are not properly correlated by the original obliquity factor.

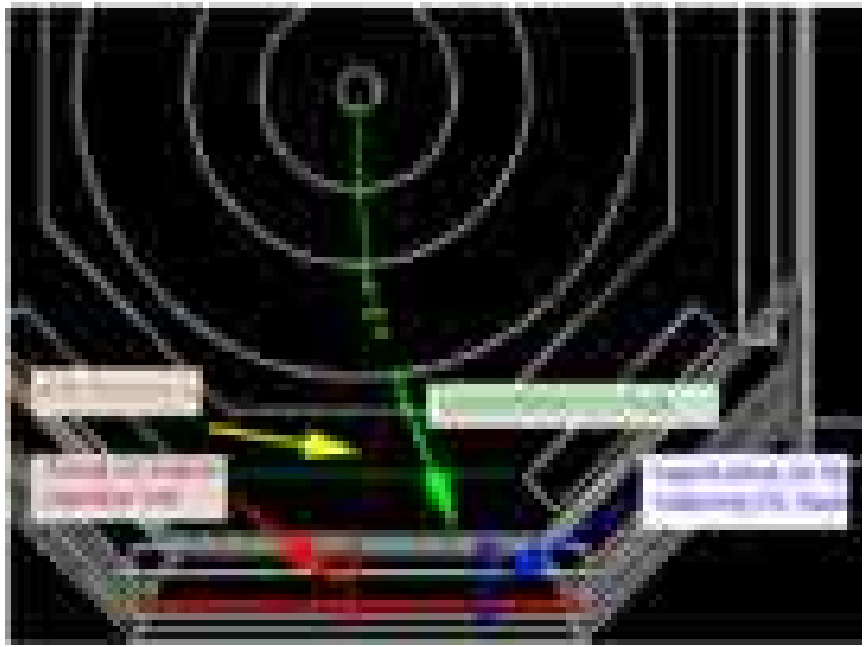

**Figure 5.1 – 3.5GeV muon event.**

Figure 5.1 is an event display of a 3.5 GeV Monte Carlo muon. We can use it to illustrate the cause of the off diagonal matrix elements. The particle trajectory through the drift chamber is shown in green. The yellow boxes are minimum ionizing depositions in the Liquid Argon Calorimeter, the yellow arrow indicates some WIC strips parallel to the beam axis (the red dots) and the red lines indicate hits in the 45-degree chambers.

If the track through the drift chamber is extrapolated through the WIC from the data gathered by the DC, it follows approximately the green arrow in the Figure 5.2. Due to multiple scattering, the actual hit is much further to the left as indicated by the red arrow. A hypothetical 45-degree hit, as indicated by the blue arrow, would align better with the track trajectory. The fitter however, should prefer the actual hit because it is much better correlated with the hits in the barrel WIC as indicated by the yellow arrow. The weight matrix reduces the penalty for large deviations if the hits are correlated. It does this through its off-diagonal matrix elements.

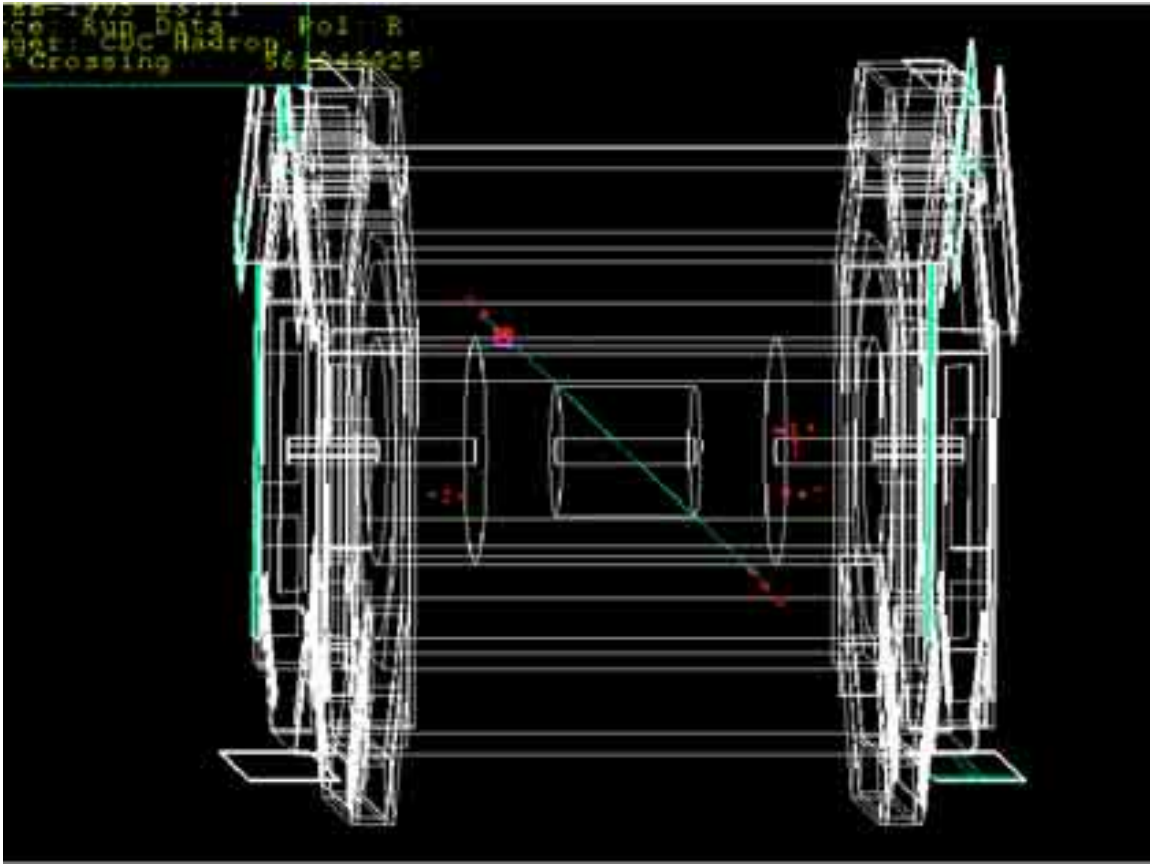

**Figure 5.2 – Display of a muon event.**

The original weight matrix didn't correctly correlate certain 45-degree chamber hits with the relevant EC and barrel hits. Figures 5.2 and 5.3 represent another 3.5 GeV muon events in the WIC. The measurement axis for the inner EC hits is vertical and the 45-degree chamber hit has a horizontal measurement axis. There was no correlation in the weight matrix because the obliquity factor is zero for two detector planes with orthogonal measurement axes. The 45-degree hit lines up well with the EC hits and poorly with the extrapolated chamber track. Without the associations being included the 45-degree hits would tend to pull the WIC track off of the correct trajectory in the direction of the multiple scatter.

A new obliquity factor with fewer simplifying assumptions was calculated, which correctly incorporated these correlations:

$$\begin{aligned}
&\left[(\hat{\alpha}_1 \bullet \hat{m}_1)(\hat{\alpha}_2 \bullet \hat{m}_2)\cos\Delta\phi / \left((\hat{t} \bullet \hat{n}_1)(\hat{t} \bullet \hat{n}_2)\right)\right] \\
&+\left[(\hat{\beta}_1 \bullet \hat{m}_1)(\hat{\beta}_2 \bullet \hat{m}_2)\cos\Delta\phi\right] \\
&+\left[(\hat{\alpha}_1 \bullet \hat{m}_1)(\hat{\beta}_2 \bullet \hat{m}_2)\sin\Delta\phi / (\hat{t} \bullet \hat{n}_1)\right] \\
&-\left[(\hat{\alpha}_2 \bullet \hat{m}_2)(\hat{\beta}_1 \bullet \hat{m}_1)\sin\Delta\phi / (\hat{t} \bullet \hat{n}_2)\right]
\end{aligned} \qquad 5.3$$

Where:

$$\hat{\alpha}_1 = \frac{(\hat{n}_1 \bullet \hat{t})\hat{n}_1 - \hat{t}}{|\hat{t} \times \hat{n}_1|}$$

$$\hat{\alpha}_2 = \frac{(\hat{n}_2 \bullet \hat{t})\hat{n}_2 - \hat{t}}{|\hat{t} \times \hat{n}_2|}$$

$$\hat{\beta}_1 = \frac{(\hat{t} \times \hat{n}_1)}{|\hat{t} \times \hat{n}_1|}$$

$$\hat{\beta}_2 = \frac{(\hat{t} \times \hat{n}_2)}{|\hat{t} \times \hat{n}_2|}$$

$$\cos\Delta\phi = \frac{\hat{n}_1 \bullet \hat{n}_2 - (\hat{t} \bullet \hat{n}_1)(\hat{t} \bullet \hat{n}_2)}{|\hat{t} \times \hat{n}_1||\hat{t} \times \hat{n}_2|}$$

$$\sin\Delta\phi = \frac{\hat{t} \bullet (\hat{n}_2 \times \hat{n}_1)}{|\hat{t} \times \hat{n}_1||\hat{t} \times \hat{n}_2|}$$

$$5.4$$

## 5.2 – New Obliquity Factor Calculation

Multiple scattering effects are included in the calculation of the inverse of the weight matrix by estimating the effect of multiple scattering on residuals in the measurement axis of each detector plane. A muon moves through a block of matter a distance Δz in the direction of the incident track in Figure 5.4. When the particle exits the block, it has been displaced due to multiple scattering by an impact parameter δ and deflected through an angle α. Its new direction is represented as $\hat{t}'$.

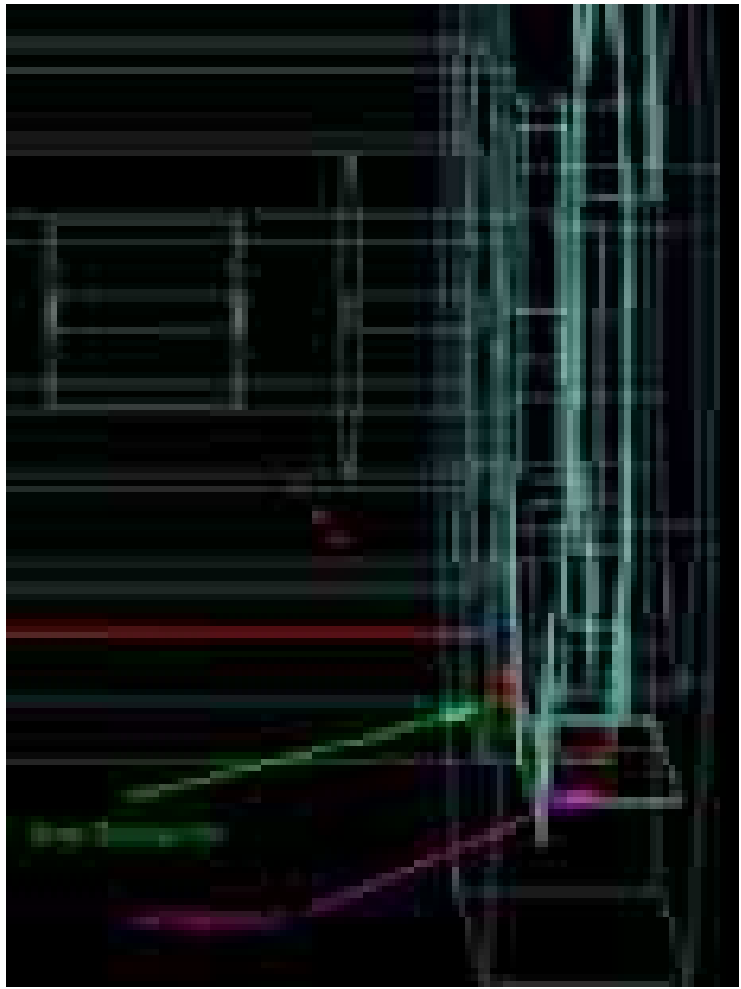

**Figure 5.3 – Muon event showing inner endcap and 45DC hits.**

Fermi calculated the joint probability density for α and δ to be:

$$\frac{12}{\pi \theta_s^4} e^{-\frac{4}{\theta_s^2}\left[\alpha^2 + \frac{3\delta^2}{\Delta z^2} - \frac{3\vec{\alpha}\cdot\vec{\delta}}{\Delta z^3}\right]} \quad 5.5$$

$\vec{\alpha}$ is perpendicular to $\hat{t}$ and has magnitude α. For small enough α This vector is $\hat{t}' - \hat{t}$. The vector $\vec{\delta}$ is perpendicular to $\hat{t}$ and has the magnitude δ. And $\hat{V}$ is the vector connecting the points where the projected track intersects the plane to the place where the scattered track intersects it. $\theta_s$ is a constant.

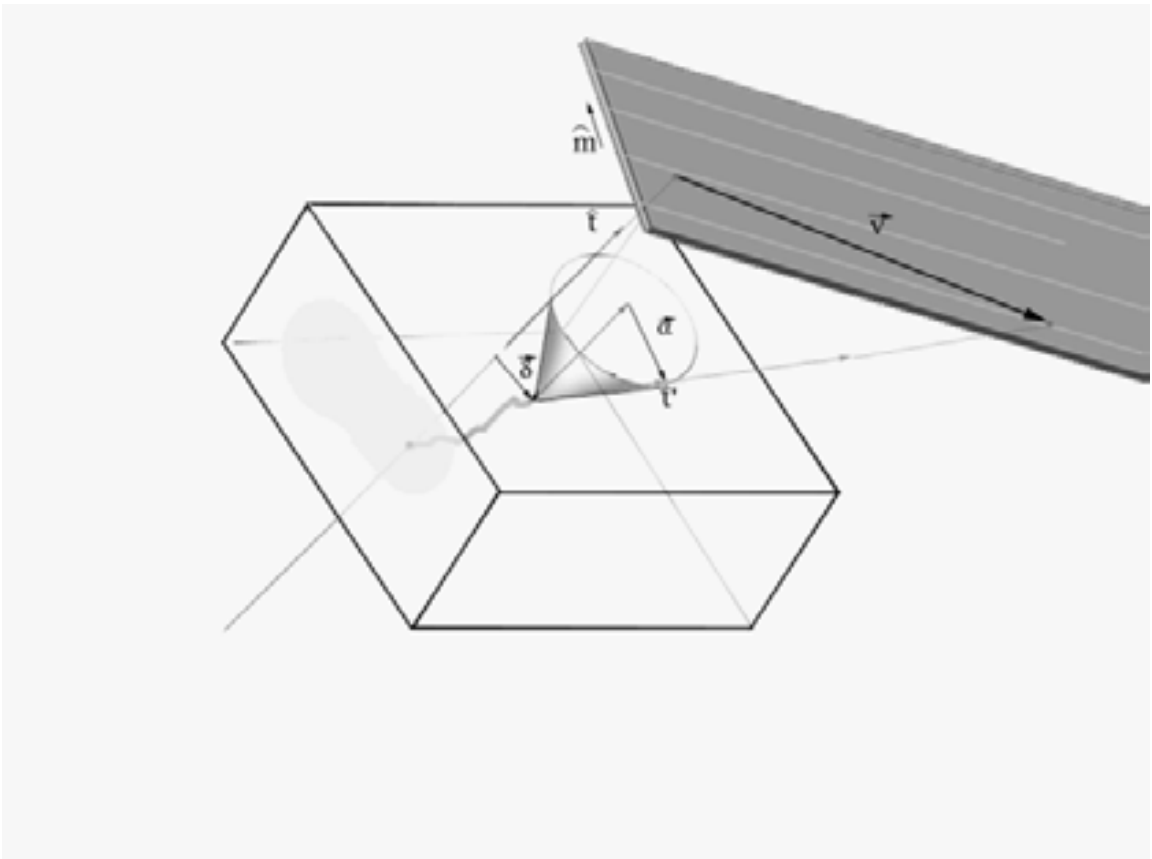

**Figure 5.4 – Movement of a particle through matter**.

The inverse of the weight matrix is:

$$W_{ij}^{-1} = \sum_{\Delta Z} \langle (\vec{V}_i \cdot \hat{m}_i)(\vec{V}_j \cdot \hat{m}_j) \rangle_{\phi_\alpha, \phi_\delta, \alpha, \delta} + \delta_{ij}\sigma_i^2 \qquad 5.6$$

The summation is the part due to multiple scattering and the $\delta_{ij}$ part is due to measurement uncertainties, including the strip width and uncorrelated alignment errors. To find the average in the first term one sets up the following coordinate systems as shown in figure 5.5:

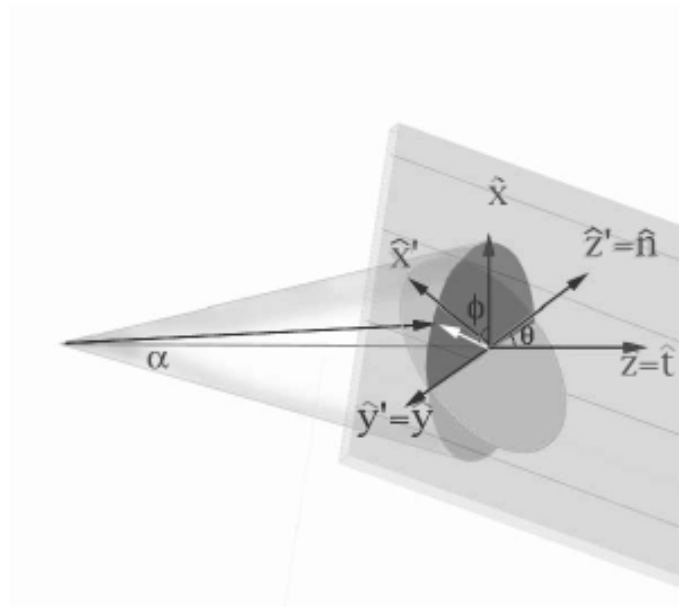

**Figure 5.5 – Reference system of particle motion relative to detector pad.**

$$\hat{z} = \hat{t}$$

$$\hat{x} = \frac{\hat{t} \times \hat{n}}{|\hat{t} \times \hat{n}|} \times \hat{t}$$

$$\hat{z}' = \hat{n} \qquad 5.7$$

$$\hat{x}' = \frac{\hat{t} \times \hat{n}}{|\hat{t} \times \hat{n}|} \times \hat{n}$$

$$\hat{y} = \hat{y}' = \frac{\hat{t} \times \hat{n}}{|\hat{t} \times \hat{n}|}$$

$\hat{n}$ is the unit vector normal to the chamber plane and $\hat{t}$ is the unit vector in the track trajectory direction. $\phi$ is the angle between the $\hat{x}$ axis and the vector $\vec{\alpha}$. S is the path length.

The coordinate axes are related by:

$$\begin{aligned} x &= x'\cos\theta + z'\sin\theta \\ y &= y' \\ z &= z'\cos\theta - x'\sin\theta \end{aligned} \qquad 5.8$$

Where $\cos\theta = \hat{t} \cdot \hat{n}$.

Points on the cone obey:

$$\begin{aligned} x &= (s+z)\tan\alpha\cos\phi \\ y &= (s+z)\tan\alpha\sin\phi \end{aligned} \qquad 5.9$$

At the detector plane, $z' = 0$, it is corresponded by $z = -x\tan\theta$.

Points of the intersection of that plane and the cone obey:

$$\begin{aligned} x_p &= x'\cos\theta = (z_p + s)\tan\alpha\cos\phi \\ y_p &= y' = (z_p + s)\tan\alpha\sin\phi \\ z_p &= -x'\sin\theta = -x_p\tan\theta \end{aligned} \qquad 5.10$$

Substituting and rearranging:

$$x_p = \frac{s \tan\alpha \cos\phi}{1 + \tan\theta \tan\alpha \cos\phi} \qquad 5.11$$

one can now obtain a new version of z-plane:

$$z_p = \frac{-s \tan\alpha \cos\phi \tan\theta}{1 + \tan\theta \tan\alpha \cos\phi} \qquad 5.12$$

and for y-plane

$$y_p = \frac{s \tan\alpha \sin\phi}{1 + \tan\theta \tan\alpha \cos\phi} \qquad 5.13$$

For small angles of deflection the second term in the denominator of each expression is small.

$$\begin{aligned} x_p &\approx s\alpha \cos\phi \\ y_p &\approx s\alpha \sin\phi \\ z_p &\approx -s\alpha \cos\phi \tan\theta \end{aligned} \qquad 5.14$$

In the primes coordinate system the vector $\vec{V}_\alpha$ (the $\alpha$ contribution to $\vec{V}$) is $(x', y', 0)$ Then one finds:

$$\vec{V}_\alpha \cdot \vec{m} = m_{x'} V_{\alpha x'} + m_{y'} V_{\alpha y'} + 0 \qquad 5.15$$

so with

$$\begin{aligned} V_{\alpha x'} &= \frac{x_p}{\cos\theta} \\ V_{\alpha y'} &= y' = y \end{aligned} \qquad 5.16$$

$$\begin{aligned} V_{\alpha x'} &= \frac{s\alpha \cos\phi}{\cos\theta} \\ V_{\alpha y'} &= s\alpha \sin\phi \end{aligned} \qquad 5.17$$

Then

$$(\vec{m} \cdot \vec{V}_\alpha) = m_{x'} V_{\alpha x'} + m_{y'} V_{\alpha y'}$$

$$= \left( \frac{\hat{t} \times \hat{n}}{|\hat{t} \times \hat{n}|} \times \hat{n} \right) \cdot \hat{m} \frac{s\alpha \cos\phi}{\cos\theta} + \frac{(\hat{t} \times \hat{n}) \cdot \hat{m}}{|\hat{t} \times \hat{n}|} (s\alpha \sin\phi) \qquad 5.18$$

The expression for $\vec{m} \cdot \vec{V}_\delta$ can be seen from the expression for $m \cdot V_\alpha$ if one recognizes that taking the small angle α and large value for s in a limit so that as α→0, sα→δ

$$(\vec{m} \cdot \vec{V}) = \left( \frac{\hat{t} \times \hat{n}}{|\hat{t} \times \hat{n}|} \times \hat{n} \right) \cdot \hat{m} \frac{s\alpha \cos\phi_\alpha + \delta \cos\phi_\delta}{\cos\theta} + \frac{(\hat{t} \times \hat{n}) \cdot \hat{m}}{|\hat{t} \times \hat{n}|} (s\alpha \sin\phi_\alpha + \delta \sin\phi_\delta) \qquad 5.19$$

In the inverse weight matrix:

$$W_{ij}^{-1} = \sum_{\Delta Z} \left\langle (\vec{V}_i \cdot \hat{m}_i)(\vec{V}_j \cdot \hat{m}_j) \right\rangle_{\phi_\alpha, \phi_\delta, \alpha, \delta} + \delta_{ij} \sigma_i^2 \qquad 5.20$$

the two detector planes i and j will have different normal directions. Therefore they will have different orientations for their axes. Their unprimed axes - $\hat{x}_i\, \hat{y}_i$ and $\hat{x}_j\, \hat{y}_j$ lie in a plane however. The angle between the two axes $\hat{y}_i$ and $\hat{y}_j$ is $\Delta\phi$.

One defines the angle by:

$$\hat{y}_1 \cdot \hat{y}_2 = \frac{\hat{t} \times \hat{n}_1}{|\hat{t} \times \hat{n}_1|} \cdot \frac{\hat{t} \times \hat{n}_2}{|\hat{t} \times \hat{n}_2|} = \cos\Delta\phi \qquad 5.21$$

along with

$$\sin\Delta\theta = \hat{x}_1 \cdot \hat{y}_2 = \frac{(\hat{t} \times \hat{n}_1) \times \hat{t}}{|\hat{t} \times \hat{n}_1|} \cdot \frac{\hat{t} \times \hat{n}_2}{|\hat{t} \times \hat{n}_2|} \qquad 5.22$$

so that one can consistently use:
$$\phi_1 + \Delta\phi = \phi_2.\qquad 5.23$$

Using:
$$(\hat{t}\times\hat{n}_1)\times\hat{t}\cdot(\hat{t}\times\hat{n}_2) = (\hat{n}_1-(\hat{n}_1\cdot\hat{t})\hat{t})\cdot(\hat{t}\times\hat{n}_2) = \hat{n}_1\cdot(\hat{t}\times\hat{n}_2)\qquad 5.24$$

One can reduce the expression for $\sin\Delta\phi$ to:
$$\sin\Delta\phi = \bar{x}_1\cdot\bar{y}_2 = \frac{t\cdot(\hat{n}_2\times\hat{n}_1)}{\sin\theta_1\sin\theta_2}\qquad 5.25$$

and using:
$$(\hat{t}\times\hat{n}_1)\cdot(\hat{t}\times\hat{n}_2) = (\hat{t}\cdot\hat{t})(\hat{n}_1\cdot\hat{n}_2)-(\hat{t}\cdot\hat{n}_1)(\hat{t}\cdot\hat{n}_2)$$
$$\cos\Delta\phi = \frac{(\hat{n}_1\cdot\hat{n}_2)-(\hat{t}\cdot\hat{n}_1)(\hat{t}\cdot\hat{n}_2)}{\sin\theta_1\sin\theta_2}\qquad 5.26$$

Now one has:
$$\phi_{\alpha_j} = \phi_{\alpha_i} + \Delta\phi$$
$$\phi_{\delta_j} = \phi_{\delta_i} + \Delta\phi\qquad 5.27$$

One defines:
$$F_1(\hat{t},\hat{n},\hat{m}) = \left(\frac{(\hat{t}\times\hat{n})\times\hat{n}}{|\hat{t}\times\hat{n}||\hat{t}\cdot\hat{n}|}\right)\cdot\hat{m}$$
$$F_2(\hat{t},\hat{n},\hat{m}) = \left(\frac{(\hat{t}\times\hat{n})}{|\hat{t}\times\hat{n}|}\right)\cdot\hat{m}\qquad 5.28$$

One now has:
$$W_{ij}^{-1} = \delta_{ij}\sigma_i^2 +$$
$$\sum_{\Delta Z}\Big\langle\big(F_1(\hat{t}_1,\hat{n}_1,\hat{m}_1)(s_1\alpha\cos\phi_\alpha+\delta\cos\phi_\delta)+F_2(\hat{t}_1,\hat{n}_1,\hat{m}_1)(s_1\alpha\cos\phi_\alpha+\delta\cos\phi_\delta)\big)$$
$$*\big(F_1(\hat{t}_2,\hat{n}_2,\hat{m}_2)(s_2\alpha\cos(\phi_\alpha+\Delta\phi)+\delta\cos(\phi_\delta+\Delta\phi))+F_2(\hat{t}_2,\hat{n}_2,\hat{m}_2)(s_2\alpha\cos(\phi_\alpha+\Delta\phi)+\delta\cos(\phi_\delta+\Delta\phi))\big)\Big\rangle\qquad 5.29$$

And using the identities:

$$\cos(\phi_\alpha + \Delta\phi) = \cos\phi_\alpha \cos\Delta\phi - \sin\phi_\alpha \sin\Delta\phi$$
$$\sin(\phi_\alpha + \Delta\phi) = \cos\phi_\alpha \sin\Delta\phi + \sin\phi_\alpha \cos\Delta\phi \qquad 5.30$$

and transforming from $\alpha, \delta, \phi_\alpha, \phi_\delta$ to $\alpha_x, \delta_x, \alpha_y, \delta_y$ and performing the integrations:

$$W_{ij}^{-1} = \delta_{ij}\sigma_i^2 + \sum_{\Delta Z} \Lambda\left(\langle\alpha^2\rangle s_1 s_2 + \langle\delta^2\rangle + \langle\alpha\delta\rangle(s_1 + s_2)\right) \qquad 5.31$$

where
$$\Lambda = F_{11}F_{12}\cos\Delta\phi - F_{12}F_{21}\sin\Delta\phi + F_{11}F_{22}\sin\Delta\phi + F_{21}F_{22}\cos\Delta\phi \qquad 5.32$$

and:

$$\langle\alpha^2\rangle = A^2 \Delta z / X$$
$$\langle\alpha\delta\rangle = \tfrac{1}{2} A^2 \Delta z^2 / X \qquad 5.33$$
$$\langle\delta^2\rangle = \tfrac{1}{3} A^2 \Delta z^3 / X$$

$$F_1(\hat{t},\hat{n},\hat{m}) = \left(\frac{(\hat{t}\times\hat{n})\times\hat{n}}{|\hat{t}\times\hat{n}||\hat{t}\cdot\hat{n}|}\right)\cdot\hat{m} = \frac{-\hat{t}\cdot\hat{m}}{|\hat{t}\times\hat{n}||\hat{t}\cdot\hat{n}|} \qquad 5.34$$

$$F_2(\hat{t},\hat{n},\hat{m}) = \left(\frac{(\hat{t}\times\hat{n})}{|\hat{t}\times\hat{n}|}\right)\cdot\hat{m} = \left(\frac{(\hat{n}\times\hat{m})}{|\hat{t}\times\hat{n}|}\right)\cdot\hat{t} \qquad 5.35$$

The results for the obliquity factor $\Lambda$ correspond to the value quoted in equation 5.3. The $\alpha^2$ term gives the dominant contribution to the inverse weight matrix the new form:

$$W_{ij}^{-1} = \delta_{ij}\sigma_i^2 + (14\text{MeV}^2)\times \int_0^{\min(s_i,s_j)} \Lambda ds' \frac{(s_i - s')(s_j - s')}{X_{rad}(s')p(s')^2} \qquad 5.36$$

**Chapter 6 – Results and Conclusions**

The 45DC geometry went through three major revisions and expansions to arrive at the completed geometry definition amounting to almost 6000 data words. It was very labor intensive to specify, requiring the creation of specialized code to reduce the labor. In the end, about 15,000 lines of FORTRAN code were written in connection with the 45DC commissioning since early 1995. This included the revised subsystem code, the programs used to create and test the fitter geometry and its pointers, and the programs used to create and test the fitter geometry and its pointers, and the programs used to create and test the fitter geometry and its pointers, and the programs that test the fitter locator routines. More than fifty routines and data files were written new or revised for the WIC code.

Figure 6.1 shows the Monte Carlo acceptance without background in cos $\theta$ and $\phi$ for high-energy muons of the WIC prior to the commissioning work. This figure shows a scatter plot of the muons that were accepted by the fitter. The void shown in the right side of the figure is the 45DC region. The horizontal voidcs correspond to a lowering of acceptance because the joints of the octants of the WIC are interfering with the fitting process. After the addition of the 45DC and fitter changes, a noticeable increase in the number of muons that were accepted is noted as seen in Figure 6.2. The remaining loss of efficiency is due partly to the complex geometry of the 45DC.

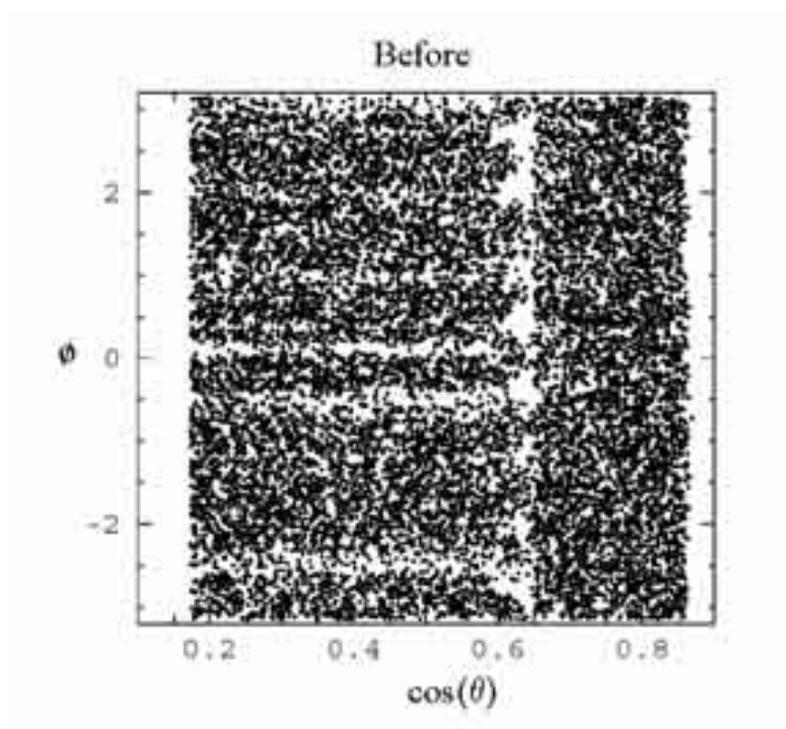

**Figure 6.1- Acceptance prior to the addition of the 45DC.**

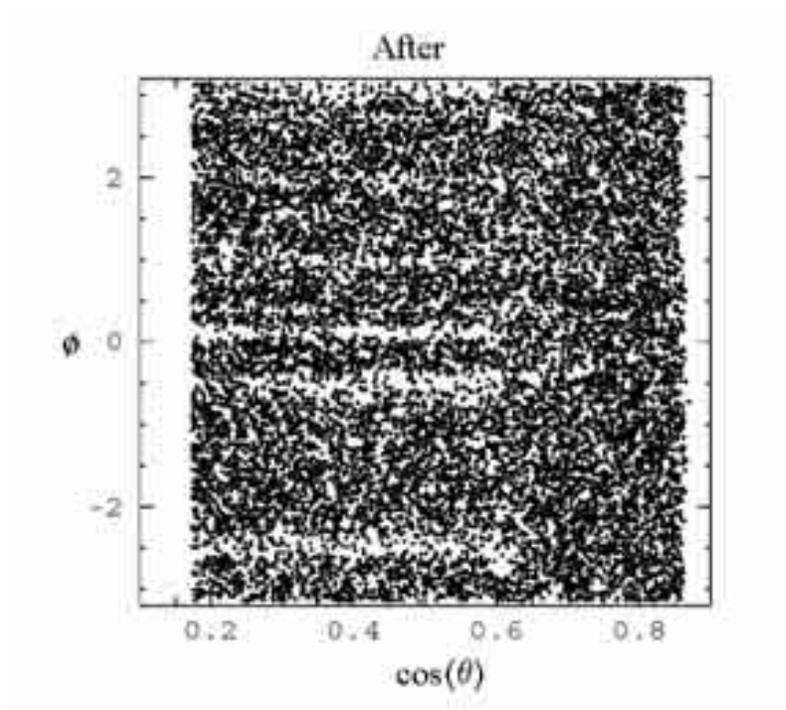

**Figure 6.2 – Muon acceptance after the addition of the 45DC.**

Figure 6.3 also illustrated the improvements by the 45DC commissioning. The graph is efficiency as a function of $\cos\theta$. It is a plot of the unimproved 45DC overlaid a plot with the 45DC corrected geometry.

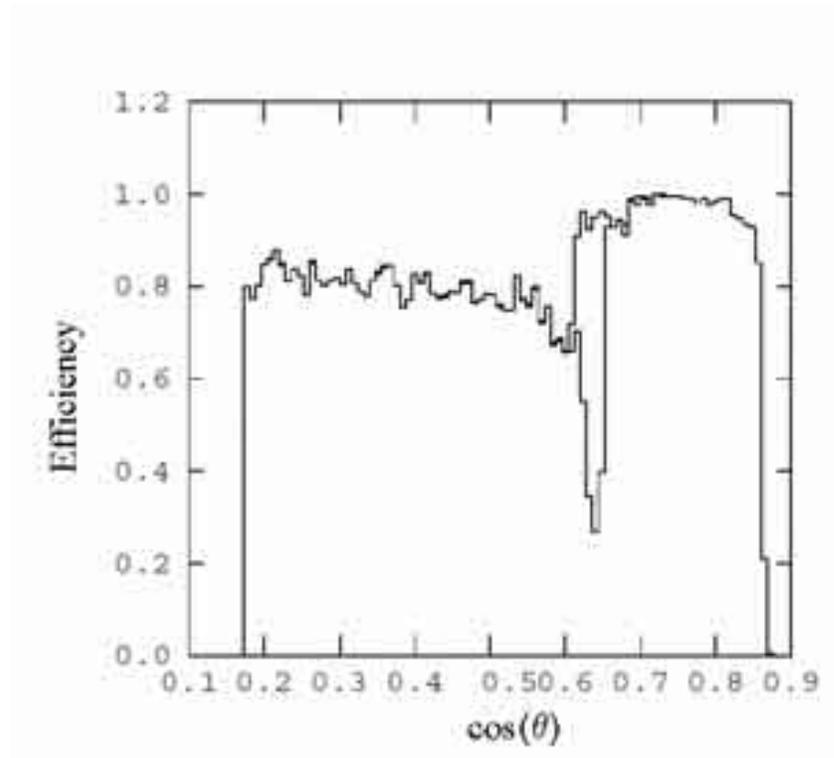

**Figure 6.3 – Muon ID efficiency overlay of the WIC for the pre/post commissioning of the 45DC.**

New and old versions of the obliquity factor are compared in Figure 6.4a. The obliquity factor in Figure 6.4a is given for a pair of planes in the octant 2 barrel (top of the WIC). $\phi$ is held constant at 22.5° from the vertical and $\theta$ varies from 35 to 145 degrees. In Figure 6.4b is the same as 6.4a except $\phi$ is held constant at the vertical. Figure 6.4c uses the planes in the inner endcap and the 45DC as in Figure 5.3. $\theta$ is held constant at 90 degrees and $\phi$ runs from 60 to 110 degrees. Figure 6.4d uses two planes of transverse strips using a constant $\phi$ held vertical and $\theta$ runs from 45 to 135 degrees.

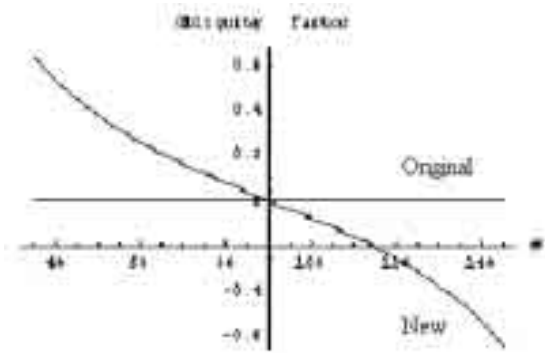
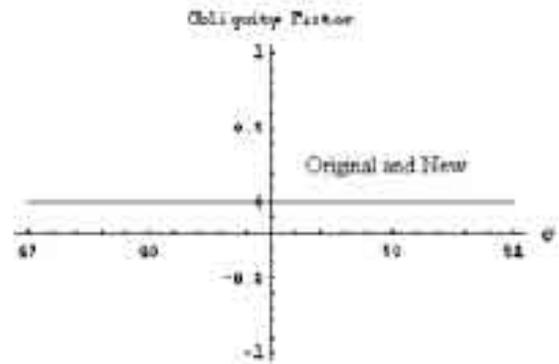

Figure 6.4a                                  Figure 6.4b

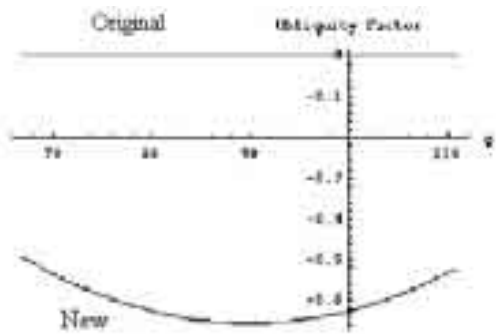
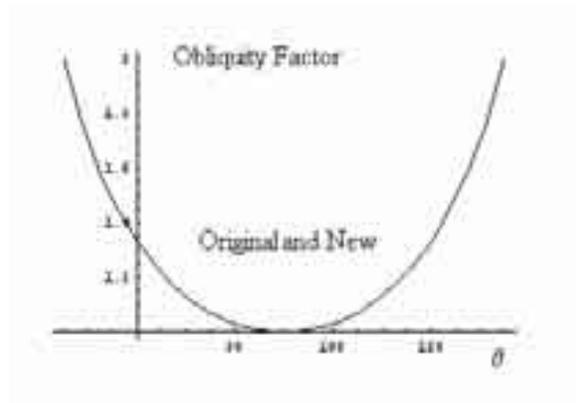

Figure 6.4c                                  Figure 6.4d

**Figures 6.4a-d are overlay plots of the original and new obliquity factors. See text for details.**

Tracking improvement caused by the new obliquity factor was minimal. The improvement due to the addition of new planes was dramatic however. It greatly improved the number of fit parameters found used in the muons fits. This is shown in Figure 6.5. The red bars represent the number of fit parameters before the addition of the 45DC. The blue bars are for the muons after the addition (passing and failing muon identification). The figure includes all events, passing or failing the muon selection.

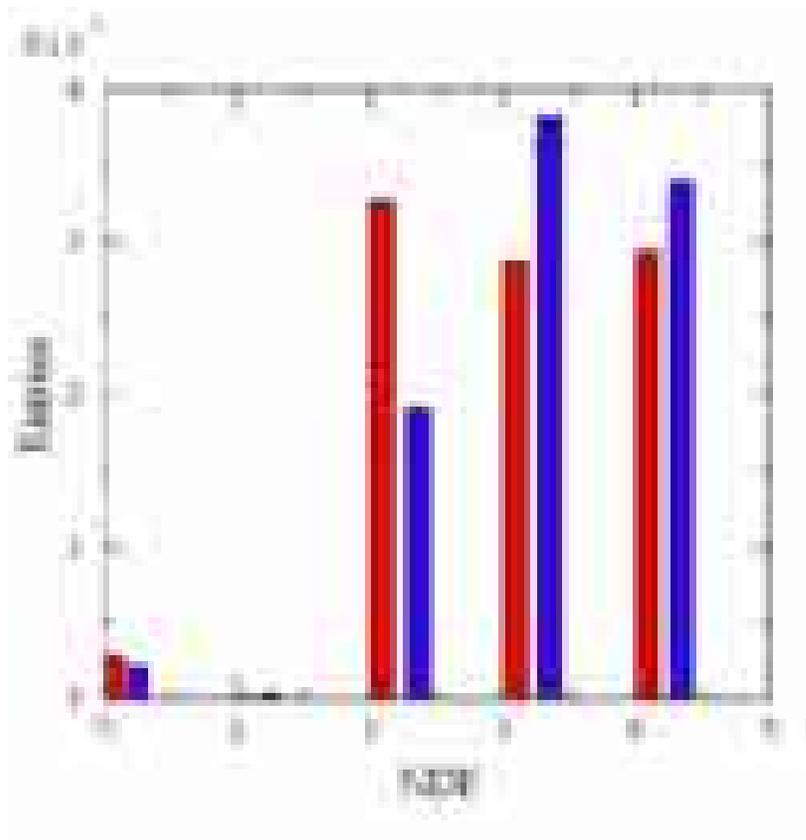

**Figure 6.5 – Number of fit parameters for all events.**

At the time this task was undertaken, the experiment was much more statistics limited than systematics limited. By the time the commissioning task was drawing to completion this was no longer quite as true. It became clear that the additional potential for physics provided by the 45DC was limited because of the systematic uncertainties associated with the strange acceptance, background muons, hot channels, uncertainties in the magnetic field in the vicinity of the chambers, and skepticism about the quality of the WIC fitter swimmer and background in the 45DC for a real event.

The inclusion of the 45DC was completed. An improved version of the WIC fitter was completed so multiple scattering can be taken into account in the 45DC. And these improvements do increase the efficiency for muon identification.

# List of References

.

# Vita

Son of a Californian Coast Guard officer and a Southern Belle from Alabama, Vance Onno Eschenburg was born in Fairfield, California. He called many places his home during his early years: San Rafael, California; Virginia Beach, Virginia; Slidell, Louisiana; Mobile, Alabama; Nimitz Hill, Guam; and Panama City, Panama. He later graduated with a Bachelor of Arts degree in Physics from the University of California at Berkeley in May, 1995. During that time, he was an active member of the Alpha Chapter of the Alpha Kappa Lambda fraternity.

Vance returned to the South to study physics at the University of Mississippi. There, he became the student of Dr. Robert Kroeger and a member of the research group studying high-energy physics. He also actively taught labs and tutored students at many levels. He is currently spending his second tour of service at the Stanford Linear Accelerator Center in Palo Alto, California.

Vance's interests are many. During high school, we was co-captain of the Quiz Bowl team and studied classical piano. Currently, he tends to his bonsai trees and his family's web site and scuba dives with his brother whenever he is near the Gulf of Mexico.

Mr. Eschenburg plans to continue his studies in high-energy physics as a member of the BaBar collaboration. He intends to obtain his Ph.D. with the University of Mississippi and then continue his work in research.